\documentclass[12pt]{article}
\usepackage{graphicx}
\usepackage{natbib} %comment out if you do not have the package
\usepackage{url} % not crucial - just used below for the URL 
\usepackage{amsmath}
\usepackage{amssymb}
\usepackage{subcaption}
\usepackage{booktabs}
\usepackage{multirow}
\usepackage{comment}
\usepackage{hyperref}
\usepackage{xcolor}
\usepackage{algorithm}
\usepackage{algpseudocode}
\usepackage{stmaryrd}
\usepackage{bbm}

\usepackage{tikz}

\usetikzlibrary{arrows.meta, positioning, fit, backgrounds, decorations.pathreplacing, patterns}
 
\definecolor{colH}{RGB}{180, 60, 60}       % historical cohort: red
\definecolor{colC}{RGB}{120, 120, 120}     % current cohort: grey
\definecolor{colU}{RGB}{80, 140, 80}       % U_k: green
\definecolor{colS}{RGB}{80, 110, 170}      % S_k: blue
\definecolor{colV}{RGB}{200, 130, 50}      % V: orange
\definecolor{colCox}{RGB}{140, 60, 160}    % Cox block: purple
\definecolor{colRank}{RGB}{50, 130, 160}   % Ranking output: teal
\definecolor{bgblock}{RGB}{245, 245, 250}  % light background
\definecolor{colW}{RGB}{80, 110, 170}   % as col S

\newcommand{\bm}[1]{\mathbf{#1}}
\newcommand{\bcal}[1]{\boldsymbol{\mathcal{#1}}}

\newcommand{\blind}{0}

\begin{document}

\def\spacingset#1{\renewcommand{\baselinestretch}%
{#1}\small\normalsize} \spacingset{1}

%%%%%%%%%%%%%%%%%%%%%%%%%%%%%%%%%%%%%%%%%%%%%%%%%%%%%%%%%%%%%%%%%%%%%%%%%%%%%%

\if0\blind
{
  \title{\bf PAR2COX: Survival-Informed Tensor Decomposition for Phenotyping and Risk Prediction from Irregular Longitudinal Data}
\author{}
\date{}
  \author{Mariafrancesca Patalano\textsuperscript{1}\footnote{Corresponding author. Email: \texttt{mariafrancesca.patalano@phd.unipd.it}\\
  %This manuscript is currently under review.
  Department of Statistical Sciences, University of Padua, Via Cesare Battisti 241, 35121 Padova, Italy.
  }, 
  Elif Konyar\textsuperscript{2}, Kamran Paynabar\textsuperscript{3}
    \hspace{.2cm}\\[1em]
    \textsuperscript{1}{Department of Statistical Sciences, University of Padua} \\
\textsuperscript{3}{School of Industrial and Systems Engineering, Georgia Institute of Technology}\\
\textsuperscript{2}{Department of Systems Engineering, The University of Texas at Dallas}
    }
    
  \maketitle
} \fi

\if1\blind
{
  \bigskip
  \bigskip
  \bigskip
  \begin{center}
    {\LARGE\bf Title}
\end{center}
  \medskip
} \fi

\bigskip
\begin{abstract}
Accurate risk prediction is crucial for clinical-decision making, intervention planning, treatment and transplant allocation. However, longitudinal clinical data are often irregular and subject to censoring. 
We propose PAR2COX, a joint framework that integrates PARAFAC2 decomposition with Cox proportional hazards model, using patient-specific latent factors as covariates in the likelihood. The proposed alternating optimization framework jointly estimates phenotypes and survival parameters, enabling survival-guided representation learning. PAR2COX accommodates both historical patients with observed outcomes and current patients whose outcomes remain unknown. 
Numerical experiments and a case study based on MIMIC-IV data demonstrate improved risk stratification compared with existing approaches, highlighting the value of survival-informed phenotype learning.

\end{abstract}

\noindent%
\vfill

\newpage
\spacingset{1.5} 
\section{Introduction}

Timely and accurate risk prediction is crucial for supporting clinical decision-making and prioritizing patients for interventions, treatment strategies, and transplant allocation. Risk stratification is particularly valuable in resource-constrained settings, such as intensive care units (ICU) where patients may deteriorate rapidly and care teams must prioritize interventions under uncertainty. 
Survival models provide a natural framework for this task because they explicitly account for time-to-event outcomes and censored observations, which commonly arise when patients are discharged, transferred, or remain event-free during follow-up \citep{cox1972,klein2003survival}. For example, in critical care settings such as sepsis management, where clinical status can evolve rapidly and deterioration may occur over short time horizons, survival modeling enables patients to be stratified according to their relative risk of adverse outcomes while accounting for differences in follow-up time and censoring \citep{singer2016third, evans2021surviving}. Such risk stratification can support dynamic patient monitoring, prioritization of limited ICU resources, and early identification of individuals at elevated near-term risk of mortality or other adverse outcomes.

Electronic health records (EHR) offer a rich source of longitudinal clinical information for survival risk prediction. Modern EHR systems routinely collect laboratory measurements, vital signs, medications, diagnoses, procedures, and demographic information, providing a high-dimensional (HD) view of patient health over time \citep{jensen2012mining, shickel2017deep}. The growing availability of such large-scale EHR data has given rise to computational phenotyping, which uses data-driven methods to extract groups of co-occurring clinical descriptors, disease characteristics, and temporal patterns that define clinically meaningful patient subgroups \citep{hripcsak2013next,shivade2014review}. 
As a result, computational phenotyping can support the interpretation of heterogeneous disease trajectories and provide compact representations for downstream predictive modeling. 
However, modeling EHR data remains challenging due to irregular patient follow-up and unevenly spaced measurements. 
Tensor decomposition methods are particularly well suited for this setting because they can represent multi-way clinical data, such as patients, visits, and clinical features, in a low-dimensional and interpretable form \citep{koldabader2009}. Among tensor decompositions, PARAFAC2 \citep{harshman1972parafac2} is especially attractive for EHR analysis because it can directly model irregular tensors in which the temporal mode varies across patients, allowing phenotype discovery without forcing all patients onto a common visit grid. Thus, PARAFAC2-based phenotyping offers a natural representation learning framework for irregular longitudinal EHR data while preserving temporal and feature-level structure relevant to downstream survival prediction. 

Existing tensor-based phenotyping approaches are commonly implemented as two-step procedures, in which phenotypes are first extracted from the EHR data and the resulting low-dimensional patient representations are then used in a separate downstream prediction model (see, e.g., \citealp{perros2019,copa,zhao2019}). This separation can limit predictive performance because the factorization objective is not informed by time-to-event outcomes and censoring information during representation learning. Consequently, phenotypes estimated from unsupervised PARAFAC2-based methods primarily explain variation in the observed longitudinal measurements, while event times and censoring information are considered only after the representation has already been estimated. 
In constrast, recent semi-supervised PARAFAC2 formulations demonstrate that label information can be incorporated into the factorization objective to guide phenotype discovery and improve prediction \citep{sspa}. However, these formulations are designed for classification settings with binary outcome labels and do not directly accommodate censored time-to-event data. Consequently, the problem of jointly learning phenotypes from irregular longitudinal EHR data while optimizing survival risk prediction remains insufficiently addressed.

In this article, we propose PAR2COX, a joint modeling framework that integrates PARAFAC2 decomposition with Cox proportional hazards regression for survival risk prediction from irregular longitudinal EHR data. The proposed framework estimates latent phenotypes and survival model parameters within a unified objective, in which patient-specific latent factor weights obtained from the PARAFAC2 decomposition act as covariates in the Cox partial likelihood. This formulation uses labeled survival data from a historical cohort to guide phenotype estimation while incorporating unlabeled patients from a current cohort during model fitting, enabling individualized risk scoring for both cohorts. 

The main contributions of this work are threefold: 
\begin{enumerate}
    \item We introduce a survival-aware PARAFAC2 formulation for irregular longitudinal EHR tensors by linking patient-specific latent weights to the Cox partial likelihood. This formulation enables the decomposition to learn phenotypes that are informed by censored time-to-event outcomes rather than by reconstruction alone. 
    \item We develop an alternating optimization algorithm with updates based on the alternating direction method of multipliers (ADMM) to estimate the tensor factors and Cox regression parameters under the proposed joint objective. The proposed optimization framework flexibly accommodates constraints, such as sparsity and temporal smoothness, to incorporate prior knowledge and enhance the interpretability of the extracted phenotypes.
    \item We develop a flexible framework that accommodates both historical-only and mixed historical–current cohort settings, enabling phenotype estimation and survival risk prediction by leveraging all available patient information, including data from patients whose outcomes remain unknown at the time of analysis.
\end{enumerate} 

The remainder of the paper is organized as follows. Section \ref{sec:literature} reviews the related literature, and Section \ref{sec:prel} introduces the notation. Section \ref{sec:methodology} presents the proposed PAR2COX approach, and Section \ref{sec:sim} evaluates its performance through simulation studies. Section \ref{sec:casestudy} presents a real-world application of PAR2COX using MIMIC-IV data, and Section \ref{sec:conclusion} concludes the paper.

\section{Literature Review}\label{sec:literature}
Advances in clinical data collection have led to the acquisition of increasingly HD data, driving the need for dimensionality reduction techniques capable of extracting interpretable and meaningful patterns. 
Matrix and tensor decomposition approaches have emerged as powerful tools for analyzing complex biomedical datasets, such as gene expression measurements and EHR. 
Among matrix factorization methods, principal component analysis (PCA) \citep{jolliffe} and non-negative matrix factorization (NMF) \citep{nmf} have been widely used in genomic studies to unveil gene expression patterns and classify patients into clinically meaningful groups (see, e.g., \citealp{brunet2004, Shen2006pca, Jia2015GeneRanking}). To improve mortality prediction in ICU, \cite{Chao2018ICUMortality} proposed a supervised extension of NMF in which the factorization and a logistic regression classifier are estimated through a joint objective function. In the context of survival analysis, \cite{nmfcox} developed a unified optimization framework that jointly minimizes a reconstruction term and a Cox partial likelihood term. 

Despite their effectiveness, matrix factorization techniques are not directly applicable to higher-order tensors.
To address this limitation, tensor decomposition methods have been developed to learn low-dimensional representations from multi-way data. Among tensor decomposition methods, CANDECOMP/PARAFAC (CP) and PARAFAC2 (see Section \ref{sec:prel}) have been widely used to extract meaningful and interpretable latent patterns from complex biomedical datasets. 
With the aim of learning clinically meaningful phenotypes, \cite{marble} introduced a non-negative CP decomposition, incorporating sparsity constraints to encourage solutions with a limited number of active components. Subsequently, this approach was extended in \citet{rubik} by incorporating guidance constraints to leverage medical domain knowledge, and pairwise constraints to promote distinct, non-overlapping phenotypes. Additional extensions have been proposed in \cite{He2019sgranite} and \cite{zhao2019}, which further incorporate structured constraints for phenotype discovery. Recently, \cite{chen2025} introduced a Cox model for matrix-variate data, where the regression coefficient matrix is decomposed using CP. 
However, CP is directly applicable to regular tensors only. 
Extensions such as CP-WOPT \citep{cpwopt} and DTW-CP \citep{dtwcp} can accommodate irregular data through dedicated preprocessing steps. Specifically, CP-WOPT was introduced to handle tensors with missing entries and can be tailored for irregular tensors via padding, recovering a regular tensor structure. DTW-CP addresses irregularity by applying dynamic time warping and by computing patient-to-pairwise distances for each feature. The resulting distance tensor is then decomposed via CP, thus operating on a preprocessed data representation rather than the raw data tensor.
In contrast, PARAFAC2 has been introduced to directly decompose irregular tensors, where one mode varies across slices, making it well suited for EHR data. 
PARAFAC2 extensions have been proposed to incorporate structured constraints, handle large, sparse, binary and missing data, and improve robustness to outliers (see, e.g., \citealp{spartan, copa, logpar, repair, atom}). In particular, \cite{copa} proposed COPA, which incorporates constraints such as temporal smoothness, sparsity and non-negativity, and has been applied to extract phenotypes and temporal patient profiles.  
With the same aim, \cite{perros2019} applied PARAFAC2 for phenotype discovery, further employing t-distributed Stochastic Neighbor Embedding (tSNE) \citep{vandermaaten08} to visualize and identify high- and low-risk patient groupings.
These methods can be regarded as two-step approaches, in which the decomposition is first applied and the resulting low-dimensional representations are subsequently used for patient grouping, patient ranking and risk prediction. 
In contrast, a joint optimization approach based on PARAFAC2 decomposition has been proposed in \cite{sspa}, where a semi-supervised learning framework is developed to incorporate label information, available for a subset of patients, to guide phenotype extraction while jointly improving mortality prediction. 
However, the method proposed in \cite{sspa} is limited to binary outcomes and cannot directly accommodate survival data, leaving open the problem of jointly integrating tensor decomposition and survival prediction for irregular clinical data.

A related but distinct line of work has developed deep learning models for survival prediction and EHR representation learning. For example, DeepSurv \citep{katzman2018deepsurv}, Cox-nnet \citep{ching2018cox}, and DeepHit \citep{lee2018deephit} learn nonlinear survival risk functions and, in some cases, relax the proportional hazards assumption, while attention-based EHR models such as RETAIN \citep{choi2016retain} learn predictive representations from longitudinal clinical histories. These approaches are highly flexible, but their learned representations are not generally organized as explicit low-dimensional phenotypes over patient, temporal, and clinical-feature modes, which is the focus of decomposition-based phenotyping methods such as PAR2COX.

\section{Preliminaries and Notation}\label{sec:prel}
This section presents the notation and relevant tensor algebra concepts used throughout the paper. In the following, scalars are denoted by lowercase letters (e.g., $x$), boldface lowercase (e.g., $\boldsymbol{x}$) denote vectors, boldface capital letters (e.g., $\bm{X}$) denote matrices, boldface calligraphic letters (e.g., $\bcal{X}$) denote tensors and calligraphic letters (e.g., $\mathcal{X}$) denote sets. In a $d-$way regular tensor $\bcal{X} \in \mathbb{R}^{I_1 \times I_2 \times \dots \times I_d}$, the dimension of each mode $i$ is $I_i$ $(i = 1, \dots, d)$. The mode-$j$ matricization of $\bcal{X}$, denoted by $\bm{X}_{(j)} \in \mathbb{R}^{I_j \times I_{-j}}$, where $I_{-j} = \prod_{i \in [d]-\{j\}} I_i$, is the matrix whose columns are the mode-$j$ fibers of $\bcal{X}$. The mode-$j$ product of a tensor $\bcal{X}$ with a matrix $\bm{Y}$ is denoted by $\bcal{X} \times_j \bm{Y}$. The mode-$j$ product can also be expressed in terms of unfolded tensors (i.e., $\bm{Y}\bm{X}_{(j)}$). In addition, the trace of a matrix $\bm{X}$ is expressed as Tr$(\bm{X})$, and $\| \cdot \|_2$ and $\| \cdot \|_F$ denote the Euclidean and Frobenius norms, respectively. The Kronecker product of two matrices $\bm{X} \in \mathbb{R}^{K \times L}$, $\bm{Y} \in \mathbb{R}^{M \times N}$ is denoted by $\bm{X} \otimes \bm{Y} \in \mathbb{R} ^{KM \times LN}$. If $L = N$, the Khatri-Rao product of $\bm{X}$ and $\bm{Y}$ is given by the columnwise Kronecker products, i.e., $\bm{X} \odot \bm{Y} = [\boldsymbol{x}_1 \otimes \boldsymbol{y}_1  \, \ldots \, \boldsymbol{x}_l \otimes \boldsymbol{y}_l] \in \mathbb{R}^{KM \times L}$. The Hadamard product is given by the element-wise multiplication of $\bm{X} \in \mathbb{R}^{K \times L}$ and $\bm{Y} \in \mathbb{R}^{K \times L}$ and is denoted as $\bm{X} * \bm{Y} \in \mathbb{R}^{K \times L}$. Additionally, we introduce the CP decomposition, which factorizes a tensor $\bcal{X} \in \mathbb{R}^{I_1 \times I_2 \times \dots \times I_d}$ into a finite sum of rank-one tensors; i.e., $\bcal{X} \approx \sum_{r=1}^R \boldsymbol{a}_1^{(r)} \circ  \ldots \circ \boldsymbol{a}_d^{(r)}$ with $\boldsymbol{a}_i^{(r)} \in \mathbb{R}^{I_i}$ and the symbol $\circ$ denotes the vector outer product. Equivalently, the CP decomposition is also expressed as $\bcal{X} \approx \llbracket \bm{A}_1, \ldots, \bm{A}_d \rrbracket$ with $\bm{A}_i \in \mathbb{R}^{I_i \times R}$. See \cite{koldabader2009} and \cite{kolda2025tensor} for a comprehensive overview of tensor decompositions and their applications. 

\section{Methodology}\label{sec:methodology}
In this section, we review the PARAFAC2 decomposition, the Cox model and detail the proposed PAR2COX approach, including parameter estimation, selection of tuning parameters, and computational complexity.

\subsection{PARAFAC2 Decomposition}
In this section, we review PARAFAC2 decomposition \citep{harshman1972parafac2}, which is applicable to irregular (ragged) tensors, where one mode is allowed to vary across slices. 
Consider $K$ patients, each with $I_k$ clinical visits, and $J$ features recorded at each visit. The resulting data forms an irregular tensor $\bcal{X} = \{\bm{X}_k\}_{k=1}^K$, where each slice $\bm{X}_k \in \mathbb{R}^{I_k \times J}$ corresponds to patient $k$. PARAFAC2 decomposes each slice as
\begin{equation}\label{eq:parafac2}
\bm{X}_k \approx \bm{U}_k \bm{S}_k \bm{V}^\top
\end{equation}
where $\bm{U}_k \in \mathbb{R}^{I_k \times R}$, $\bm{S}_k \in \mathbb{R}^{R \times R}$ is diagonal, $\bm{V} \in \mathbb{R}^{J \times R}$, and $R$ is the number of latent factors (i.e., the rank of the decomposition), also interpreted as the number of phenotypes. In Eq.\,(\ref{eq:parafac2}), $\bm{U}_k$ contains the temporal evolution of the latent factors for patient $k$, $\bm{S}_k$ encodes the patient-specific weight of each latent factor, and $\bm{V}$ maps each clinical feature $j$ to each latent factor $r$. To ensure the uniqueness of the decomposition, the constraint $\bm{U}_k^\top \bm{U}_k = \bm{\Phi}$ is imposed, where $\bm{\Phi}$ is an unknown matrix that is independent of $k$. This constraint can be equivalently expressed as $\bm{U}_k = \bm{Q}_k \bm{H}$, where $\bm{Q}_k \in \mathbb{R}^{I_k \times R}$ is constrained to be orthonormal and $\bm{H} \in \mathbb{R}^{R \times R}$, with $\bm{H}$ independent of $k$. Notably, the reparameterization $\bm{U}_k = \bm{Q}_k \bm{H}$ ensures that $\bm{U}_k^\top \bm{U}_k = \bm{\Phi}$, for $k = 1, \dots, K$, hold by construction as $\bm{U}_k^\top \bm{U}_k = \bm{H}^\top \bm{Q}_k^\top \bm{Q}_k \bm{H} = \bm{H}^\top \bm{H} = \bm{\Phi}$. 

\subsection{Cox Model}
The Cox proportional hazards model \citep{cox1972} is one of the most widely used methods for assessing the relationship between covariates and event times in the presence of censoring. 
Let $(t_k, \delta_k, \boldsymbol{z}_k)_{k=1}^K$ denote the observed data for $K$ individuals, 
where $t_k = \min(D_k, C_k)$ is the observed time, $D_k$ the true survival time, $C_k$ the 
censoring time, and $\delta_k$ the event indicator, with $\delta_k = 1$ if the event (e.g., death) is observed and $\delta_k = 0$ if the observation is censored. For each $k-$th individual, 
$\boldsymbol{z}_k \in \mathbb{R}^{P}$ denotes the vector of covariates. The Cox proportional 
hazards model specifies the hazard function as
\begin{equation}\label{eq:cox}
    h(t \mid \boldsymbol{z}_k) = h_0(t) \exp(\boldsymbol{z}_k^\top \boldsymbol{\gamma})
\end{equation}
for $k = 1, \ldots, K$. In Eq.\,(\ref{eq:cox}), $h_0(t)$ is an unspecified non-negative baseline hazard function and $\boldsymbol{\gamma} \in \mathbb{R}^{P}$ is the vector of regression coefficients. The Cox proportional hazards model, hereinafter referred to as the 
Cox model, is semi-parametric, combining a non-parametric baseline hazard function with 
a parametric linear predictor.
Since $h_0(\cdot)$ is unspecified, the vector of parameters $\boldsymbol{\gamma}$ cannot be estimated via standard likelihood maximization. To address this, \cite{cox1975} introduced the partial likelihood, defined as
\begin{equation}\label{eq:partlog}
    \mathcal{L}(\boldsymbol{\gamma}) = \prod_{k \in \mathcal{O}} \frac{\exp{(\boldsymbol{z}_k^\top \boldsymbol{\gamma})}}{\sum_{\ell \in \mathcal{R}(t_k)} \exp{(\boldsymbol{z}_\ell^\top \boldsymbol{\gamma})}}
\end{equation}
which depends neither on $h_0(\cdot)$ nor on the actual event times, but only on their ordering. In Eq.\,(\ref{eq:partlog}), $\mathcal{O} = \{k:\, \delta_k = 1\}$ denotes the set of indices for which an event is observed, $\mathcal{R}(t_k) = \{\ell : t_\ell \geq t_k\}$ is the risk set at time $t_k$, and ties 
among event times are assumed absent. For a comprehensive review of survival analysis, refer to \cite{klein2003survival}.

\subsection{PAR2COX}
The proposed PAR2COX approach jointly learns survival-informed patient phenotypes and a survival prediction model from irregular longitudinal data.
Specifically, it simultaneously optimizes a low-rank PARAFAC2 decomposition and a Cox proportional hazards objective, such that the latent patient representations are guided by survival outcomes. The resulting latent factors capture clinically interpretable patient subgroups and serve as inputs to the Cox model, while the Cox survival objective simultaneously guides the learning of these latent representations to preserve survival-related information and the relative risk ordering among subjects.

We consider $K$ patients, each having measurements collected in $\bm{X}_k \in \mathbb{R}^{I_k \times J}$, where $I_k$ denotes the number of visits and $J$ the number of clinical variables (i.e., features) recorded at each visit. Additionally, each patient $k$ is characterized by a set of time-invariant covariates, denoted as $\boldsymbol{z}_k \in \mathbb{R}^{P}$, such as sex, ethnicity, and blood type. Furthermore, the $K$ patients are partitioned into two groups: ($i$) a historical cohort, comprising patients who have either experienced the event of interest or have been right-censored, and ($ii$) a current cohort, comprising patients who are currently under observation (e.g., hospitalized) and whose outcomes remain unknown at the time of analysis. The historical and current cohorts therefore correspond to the labeled and unlabeled patient sets, respectively.
In particular, let $\{(t_k, \delta_k, \boldsymbol{z}_k)\}_{k=1}^K$ denote the survival data associated with the $K$ patients, where $t_k = \min(D_k, C_k)$ is the observed time, with $D_k$ the true survival time and $C_k$ the censoring time, and $\delta_k$ the binary event indicator. In addition, let $\mathcal{H}$ and $\mathcal{C}$ denote the historical and current cohorts, respectively, and $K = |\mathcal{H}| + |\mathcal{C}|$. 

PAR2COX jointly decomposes each slice $\bm{X}_k$ $(k = 1,\dots, K)$ using PARAFAC2, while minimizing the negative Cox log-likelihood with covariates given by $\boldsymbol{z}_k \in \mathbb{R}^{P}$ and patient-specific latent weights $\boldsymbol{w}_k \in \mathbb{R}^{R}$ obtained from the decomposition (see Figure \ref{fig:par2cox}). Therefore, the objective function is given by
\begin{multline*}
  \min_{\{\bm{U}_k\}, \{\bm{S}_k\}, \bm{V}, \boldsymbol{\beta}, \boldsymbol{\gamma}}\quad
  \sum_{k=1}^K \frac{1}{2} \| \bm{X}_k - \bm{U}_k \bm{S}_k \bm{V}^\top\|_F^2
  + \frac{\alpha}{2} \| \boldsymbol{\beta} \|_2^2 + \frac{\eta}{2} \| \bm{W}\|_F^2 \\
  - \lambda \, \frac{1}{n_\mathcal{O}} \sum_{k \in \mathcal{O}} \left[ \boldsymbol{w}_k^\top \boldsymbol{\beta}
  + \boldsymbol{z}_k^\top \boldsymbol{\gamma}
  - \log \sum_{\ell \in \mathcal{R}(t_k)} \exp \left( \boldsymbol{w}_\ell^\top \boldsymbol{\beta}
  + \boldsymbol{z}_\ell^\top \boldsymbol{\gamma}\right)\right],
\end{multline*}
where $\bm{U}_k \in \mathbb{R}^{I_k \times R}$ and $\bm{S}_k \in \mathbb{R}^{R \times R}$ are factor matrices for patient $k$, $\bm{V} \in \mathbb{R}^{J \times R}$ is shared across patients, and $R$ is the rank of the decomposition. The matrix $\bm{W} \in \mathbb{R}^{K \times R}$ contains in each row $k$ the vector $\boldsymbol{w}_k = \mathrm{diag}(\bm{S}_k)$, $\mathcal{O} = \{k \in \mathcal{H}:\,\delta_k=1\}$ denotes the set of patients for which the event of interest is observed, and $n_\mathcal{O} = |\mathcal{O}|$. In addition, $\mathcal{R}(t_k) = \{\ell \in \mathcal{H} \cup \mathcal{C} : t_\ell \geq t_k\}$ is the risk set at time $t_k$, $\boldsymbol{\beta} \in \mathbb{R}^R$ and $\boldsymbol{\gamma} \in \mathbb{R}^{P}$ are the vector of parameters associated with the latent and time-invariant covariates, respectively, and $\alpha$, $\eta$ and $\lambda$ are non-negative tuning parameters. 

To ensure the uniqueness of the decomposition, we impose the constraints $\bm{U}_k = \bm{Q}_k \bm{H}$ $(k = 1, \dots, K)$, with $\bm{Q}_k \in \mathbb{R}^{I_k \times R}$, $\bm{Q}_k^\top \bm{Q}_k = \bm{I} \in \mathbb{R}^{R \times R}$ and $\bm{H} \in \mathbb{R}^{R \times R}$ \citep{harshman1972parafac2}. 
% Generalized Constraints from COPA
We further extend the objective by introducing generalized constraints (e.g., temporal smoothness, sparsity and non-negativity) on the factor matrices.
These constraints allow us to incorporate prior knowledge and structural assumptions into the decomposition, and the resulting factors are often more interpretable and better aligned with the underlying real-world application.
Therefore, generalized constraints are introduced for $\bm{H}$, $\bm{S}_k$, $\bm{V}$, denoted as $c(\bm{H})$, $c(\bm{S}_k)$ and $c(\bm{V})$, and auxiliary variables $\overline{\bm{H}}$, $\overline{\bm{V}}$ and $\overline{\bm{S}}_k$ $(k = 1,\dots, K)$ are introduced. Hence, the proposed PAR2COX problem is formulated as follows:
\begin{multline}\label{eq:par2coxcomplete}
  \min_{\{\bm{Q}_k\}, \{\bm{S}_k\}, \bm{V}, \bm{H},\boldsymbol{\beta}, \boldsymbol{\gamma}}\quad
  \sum_{k=1}^K \frac{1}{2} \| \bm{X}_k - \bm{Q}_k \bm{H} \bm{S}_k \bm{V}^\top\|_F^2
  + \frac{\alpha}{2} \| \boldsymbol{\beta} \|_2^2 + \frac{\eta}{2} \| \bm{W}\|_F^2 \\
  - \lambda \, \frac{1}{n_\mathcal{O}} \sum_{k \in \mathcal{O}} \left[ \boldsymbol{w}_k^\top \boldsymbol{\beta}
  + \boldsymbol{z}_k^\top \boldsymbol{\gamma}
  - \log \sum_{\ell \in \mathcal{R}(t_k)} \exp \left( \boldsymbol{w}_\ell^\top \boldsymbol{\beta}
  + \boldsymbol{z}_\ell^\top \boldsymbol{\gamma}\right)\right] \\
  + c(\bm{H}) + \sum_{k=1}^K c(\bm{S}_k) + c(\bm{V})\\
  s.t. \quad \quad \bm{Q}_k^\top \bm{Q}_k = \bm{I} \quad \forall k, \overline{\bm{H}} = \bm{H}, \overline{\bm{V}} = \bm{V}, \overline{\bm{S}}_k = \bm{S}_k \quad \forall k.  \quad \quad \quad 
\end{multline}

To estimate the model parameters, we use alternating optimization with the alternating direction method of multipliers (AO-ADMM) framework \citep{huangaoadmm}. Specifically, at each outer iteration, we first update $\{\bm{Q}_k\}_{k=1}^K$. At each inner iteration, $\bm{H}$, $\bm{V}$, $\{\bm{S}_k\}_{k=1}^K$, and Cox parameters $\boldsymbol{\beta}$ and $\boldsymbol{\gamma}$ are updated cyclically by solving \eqref{eq:par2coxcomplete} for one variable while fixing the remaining ones. This procedure is repeated until convergence. In the following, we detail the update for each variable.

First, at each outer iteration, we update $\bm{Q}_k$ for $k = 1, \dots, K$. When solving for each $k$, the problem reduces to
\begin{equation}
\label{eq:solveq}
\begin{aligned}
    \min_{\bm{Q}_k} \quad
  &\frac{1}{2} \| \bm{X}_k - \bm{Q}_k \bm{H} \bm{S}_k \bm{V}^\top\|_F^2,  \\ s.t. \quad &\bm{Q}_k^\top \bm{Q}_k = \bm{I},
\end{aligned}
\end{equation}
which is an orthogonal Procrustes problem. The optimal solution is given by $\bm{Q}_k = \bm{Z}_k \bm{P}_k^\top$, where $\bm{Z}_k \in \mathbb{R}^{I_k \times R}$ and $\bm{P}_k \in \mathbb{R}^{R \times R}$ are obtained from the singular value decomposition (SVD) of $\bm{X}_k \bm{V} \bm{S}_k \bm{H}^\top$.

Subsequently, with $\{\bm{Q}_k\}_{k=1}^K$ fixed, Eq.\,\eqref{eq:par2coxcomplete} is solved for $\bm{H}$, $\bm{V}$, $\{\bm{S}_k\}_{k=1}^K$ (equiv. $\bm{W}$), $\boldsymbol{\beta}$ and $\boldsymbol{\gamma}$. Note that minimizing $\sum_{k=1}^K \frac{1}{2} \| \bm{Q}_k^\top\bm{X}_k -  \bm{H} \bm{S}_k \bm{V}^\top\|_F^2$ with respect to $\bm{H}$, $\bm{V}$ and $\{\bm{S}_k\}$ is equivalent to performing a CP decomposition of the tensor $\bcal{Y} \in \mathbb{R}^{R \times J \times K}$ with slices $\bm{Y}_k = \bm{Q}_k^\top \bm{X}_k \in \mathbb{R}^{R \times J}$ \citep{Kiers1999}. Consequently, Eq.\,\eqref{eq:par2coxcomplete} is reformulated as 
\begin{equation}
    \label{eq:par2coxtensor}
    \begin{aligned}
          &\min_{ \bm{H},\bm{V}, \bm{W}, \boldsymbol{\beta}, \boldsymbol{\gamma}}\quad
 \sum_{k=1}^K \frac{1}{2} \| \bcal{Y} - \llbracket \bm{H};\bm{V}; \bm{W} \rrbracket\|_F^2
  + \frac{\alpha}{2} \| \boldsymbol{\beta} \|_2^2 + \frac{\eta}{2} \| \bm{W}\|_F^2 \\&
  \qquad\qquad- \lambda \, \frac{1}{n_\mathcal{O}} \sum_{k \in \mathcal{O}} \left[ \boldsymbol{w}_k^\top \boldsymbol{\beta}
  + \boldsymbol{z}_k^\top \boldsymbol{\gamma}
  - \log \sum_{\ell \in \mathcal{R}(t_k)} \exp \left( \boldsymbol{w}_\ell^\top \boldsymbol{\beta}
  + \boldsymbol{z}_\ell^\top \boldsymbol{\gamma}\right)\right] \\&
  \qquad\qquad+ c(\bm{H}) +  c(\bm{W}) + c(\bm{V}),\\&
  s.t. \quad\quad \overline{\bm{H}} = \bm{H}, \overline{\bm{V}} = \bm{V}, \overline{\bm{W}} = \bm{W}, 
    \end{aligned}
\end{equation}
where $\bm{S}_k = \text{diag}(\boldsymbol{w}_k)$, with $\boldsymbol{w}_k = \bm{W}(k,:) \in \mathbb{R}^{1 \times R}$, and $\overline{\bm{W}}$ is the auxiliary variable. 

At each inner iteration $t$ of the ADMM, we first update $\bm{H}$ following 
\begin{equation}\label{eq:updateH}
\begin{aligned}
\bm{H}^t &= \left( \bm{Y}_{(1)} \,(\bm{W}^{t-1} \odot \bm{V}^{t-1})  \,+ \, {\eta}_{H}(\overline{\bm{H}}^{t-1} + \bm{\Gamma}_{H})  \right)\left( (\bm{W}^{t-1 \top}\bm{W}^{t-1} \ast  \bm{V}^{t-1\top} \bm{V}^{t-1} + {\eta}_{H}\bm{I}\right)^{\dag},\\
\overline{\bm{H}}^t &= \underset{\overline{\bm{H}}}{\text{argmin}} \,  c(\overline{\bm{H}})  + \frac{{\eta}_{H}}{2} \|\overline{\bm{H}} - \bm{H}^t + \bm{\Gamma}_{{H}}^{t-1} \|_F^2,\\
\bm{\Gamma}_{{H}}^{t} &= \bm{\Gamma}_{{H}}^{t-1} + \overline{\bm{H}}^t - \bm{H}^t,
\end{aligned}
\end{equation}
where $\bm{Y}_{(1)}$ is the mode$-1$ matricization of $\bcal{Y}$, $\dag$ denotes the Moore-Penrose pseudoinverse, $\bm{\Gamma}_H$ is the dual variable and $\eta_H$ is a positive scalar. 

Given $\{\bm{Q}_k\}_{k=1}^K$ and $\bm{H}$ fixed, we next update $\bm{V}$ at inner iteration $t$, following a similar procedure that leverages the mode$-2$ matricization of $\bcal{Y}$, yielding 
\begin{equation}\label{eq:updateV}
\begin{aligned}
\bm{V}^t &= \left( \bm{Y}_{(2)} \,(\bm{W}^{t-1} \odot \bm{H}^{t-1})  \,+ \, {\eta}_{V}(\overline{\bm{V}}^{t-1} + \bm{\Gamma}_{V})  \right)\left( (\bm{W}^{t-1 \top}\bm{W}^{t-1} \ast  \bm{H}^{t-1\top} \bm{H}^{t-1} + {\eta}_{V}\bm{I}\right)^{\dag},\\
\overline{\bm{V}}^t &= \underset{\overline{\bm{V}}}{\text{argmin}} \,  c(\overline{\bm{V}})  + \frac{{\eta}_{V}}{2} \|\overline{\bm{V}} - \bm{V}^t + \bm{\Gamma}_{{V}}^{t-1} \|_F^2,\\
\bm{\Gamma}_{{V}}^{t} &= \bm{\Gamma}_{{V}}^{t-1} + \overline{\bm{V}}^t - \bm{V}^t.
\end{aligned}
\end{equation}

Subsequently, with $\{\bm{Q}_k\}_{k=1}^K$, $\bm{H}$ and $\bm{V}$ fixed, we update $\bm{W}$ row-wise, updating each $\boldsymbol{w}_k$ according to whether patient $k$ is censored or not. 
Let $\boldsymbol{y}_k \in \mathbb{R}^{1 \times RJ}$ denote the $k-$th row of $\bm{Y}_{(3)}$. The gradient of the objective function in \eqref{eq:par2coxtensor} with respect to $\boldsymbol{w}_k$ is given by
\begin{equation}\label{eq:derivwk}
\begin{aligned}
    \nabla_{\boldsymbol{w}_k} = &-\boldsymbol{y}_k (\bm{V} \odot\bm{H}) + \boldsymbol{w}_k (\bm{V} \odot \bm{H})^\top (\bm{V} \odot \bm{H}) + \eta \boldsymbol{w}_k + \eta_W \boldsymbol{w}_k - \eta_W(\overline{\boldsymbol{w}}_k +\boldsymbol{\gamma}_{W,k})\\
    &-\frac{\lambda}{n_\mathcal{O}} \, \delta_k \, \boldsymbol{\beta}^\top + \frac{\lambda}{n_\mathcal{O}} \sum_{i:k \in \mathcal{R}(t_i)} \, \frac{\exp \left( \boldsymbol{w}_k^\top \boldsymbol{\beta}
  + \boldsymbol{z}_k^\top \boldsymbol{\gamma} \right)}{\sum_{\ell \in \mathcal{R}(t_i)} \exp \left( \boldsymbol{w}_\ell^\top \boldsymbol{\beta}
  + \boldsymbol{z}_\ell^\top \boldsymbol{\gamma}\right)} \,\boldsymbol{\beta}^\top,
\end{aligned}
\end{equation}
where $\boldsymbol{\gamma}_{W,k}$ is the $k-$th row of the dual variable $\bm{\Gamma}_W$, $\eta$ and $\eta_W$ are positive scalars. 
Setting $\nabla_{\boldsymbol{w}_k} = \boldsymbol{0}$ and solving for $\boldsymbol{w}_k$ yields the update $\boldsymbol{w}_k^t$. We solve Eq.\,(\ref{eq:derivwk}) numerically using a trust region Newton-type solver. 
Therefore, the updates of $\bm{W}$ at iteration $t$ are given by
\begin{equation}\label{eq:updateW}
\begin{aligned}
\bm{W}^t &= \big[\boldsymbol{w}_1^t; \dots; \boldsymbol{w}_K^t\big], \quad \text{where each } \boldsymbol{w}_k^t \text{ solves } \nabla_{\boldsymbol{w}_k} = \boldsymbol{0} \text{ in Eq.\eqref{eq:derivwk}}\,,\\
\overline{\bm{W}}^t &= \underset{\overline{\bm{W}}}{\text{argmin}} \,  c(\overline{\bm{W}})  + \frac{{\eta}_{W}}{2} \|\overline{\bm{W}} - \bm{W}^t + \bm{\Gamma}_{{W}}^{t-1} \|_F^2,\\
\bm{\Gamma}_{{W}}^{t} &= \bm{\Gamma}_{{W}}^{t-1} + \overline{\bm{W}}^t - \bm{W}^t.
\end{aligned}
\end{equation}
The last update involves the parameters $\boldsymbol{\beta}$ and $\boldsymbol{\gamma}$, which appear in the Cox log-likelihood. Specifically,  with $\{\bm{Q}_k\}_{k=1}^K$, $\bm{H}$, $\bm{V}$, and $\bm{W}$ fixed, the negative log-likelihood and the ridge penalty for $\boldsymbol{\beta}$ are minimized with respect to $\boldsymbol{\theta} = (\boldsymbol{\beta}^\top, \boldsymbol{\gamma}^\top) \in \mathbb{R}^{(R+P) \times 1}$ using a quasi-Newton algorithm. 

The inner ADMM iterations for updating $\bm{H}$, $\bm{V}$, $\bm{W}$ and $\boldsymbol{\theta}$ are repeated for each outer iteration $g$, and the entire procedure is repeated until convergence (i.e., $\frac{|l^g - l^{g-1}|}{l^{g-1}} < \varepsilon$, with $l^g$ the objective function value of (\ref{eq:par2coxcomplete}) at outer iteration $g$ and $\varepsilon$ the tolerance). 
The complete update procedure is summarized in Algorithm~\ref{alg:par2cox}. 

\begin{algorithm}[t]
\caption{PAR2COX (AO-ADMM)}
\label{alg:par2cox}
\begin{algorithmic}[1]
\State \textbf{Initialize:} $\{\bm{Q}_k\}$, $\bm{H}$, $\bm{V}$, $\bm{W}$, $\boldsymbol{\beta}$, $\boldsymbol{\gamma}$, auxiliary variables $\overline{\bm{H}}, \overline{\bm{V}}, \overline{\bm{W}}$, dual variables $\boldsymbol{\Gamma}_H, \boldsymbol{\Gamma}_V, \boldsymbol{\Gamma}_W$
\State $g \gets 0$, $l^0 \gets \infty$
\While{$g < g_{\max}$ \textbf{and} ${|l^g - l^{g-1}|}/{l^{g-1}} \geq \varepsilon$}
    \State $g \gets g + 1$
    \State Update $\bm{Q}_k$, for $k = 1,\dots,K$, as in Eq.~(\ref{eq:solveq})
    \State Form $\bm{Y}_k = \bm{Q}_k^\top \bm{X}_k$, for $k = 1,\dots,K$ 
    \For{$t = 1$ \textbf{to} $T_{in}$}
        \State Update $\bm{H}$, $\overline{\bm{H}}$, $\boldsymbol{\Gamma}_H$ as in Eq.~(\ref{eq:updateH})
        \State Update $\bm{V}$, $\overline{\bm{V}}$, $\boldsymbol{\Gamma}_V$ as in Eq.~(\ref{eq:updateV})
        \State Update $\bm{W}$, $\overline{\bm{W}}$, $\boldsymbol{\Gamma}_W$ as in Eq.~(\ref{eq:updateW})
        \State Update $\boldsymbol{\beta}$, $\boldsymbol{\gamma}$ via quasi-Newton minimization 
    \EndFor
    \State Compute objective value $l^g$ as in Eq.~(\ref{eq:par2coxcomplete})
\EndWhile
\end{algorithmic}
\end{algorithm}

\begin{figure}[t]
  \centering
\begin{tikzpicture}[
  font=\small,
  >=Stealth,
  every node/.style={align=center},
]

\begin{scope}[xshift=-1.0cm, yshift=-0.5cm]

\foreach \i/\h in {2/2.5, 1/2.0, 0/1.8}{
  \draw[draw=colH, fill=colH!6, thick]
    ({0.22*\i}, {-0.22*\i}) rectangle ({0.22*\i+2.5}, {-0.22*\i-\h});
}
\node at (1.0+0.25, -1.0) {$\bm{X}_k$};
 
\draw[decorate,
      decoration={brace, amplitude=5pt, raise=3pt, mirror}]
  (0.0, 0.0) -- (0.0, -1.8)
  node[midway, left=5pt, font=\footnotesize] {$I_k$};
 
\draw[decorate,
      decoration={brace, amplitude=5pt, raise=3pt}]
  (0.0, 0) -- (2.5, 0)
  node[midway, above=8pt, font=\footnotesize] {$J$};
 
\node[font=\footnotesize\bfseries, colH] at (-1.2, -0.1) {$\mathcal{H}$};
\node[font=\tiny, colH] at (1.6, -3.3) {$k \in \mathcal{H}$};

\foreach \i/\h in {1/2.8, 0/2.2}{
  \draw[draw=colC, fill=colC!5, thick, dashed]
    ({0.22*\i}, {-4.6-0.22*\i}) rectangle ({0.22*\i+2.5}, {-4.6-0.22*\i-\h});
}
\node at (1.0+0.25, -5.7) {$\bm{X}_k$};
\node[font=\footnotesize\bfseries, colC] at (-1.2, -4.7) {$\mathcal{C}$};
\node[font=\tiny, colC] at (1.6, -8) {$k \in \mathcal{C}$};

\draw[draw=colH, fill=colH!5, thick, rounded corners=4pt]
  (3.6, -1.0) rectangle (5.5, -2.2);
\node[font=\footnotesize] at (4.55, -1.6)
  {$(t_k,\,\delta_k,\boldsymbol{z}_k)$};

\draw[draw=colC, fill=colC!5, thick, dashed, rounded corners=4pt]
  (3.6, -5.2) rectangle (5.5, -6.4);
\node[font=\footnotesize, colC] at (4.55, -6.1)
  {$(t_k, \delta_k, \boldsymbol{z}_k)$\\};
 
\node at (3.25, -1.6) {$+$};
\node at (3.25, -5.9) {$+$};

\end{scope}

\foreach \i/\h in {4/2.8, 3/2.2, 2/2.5, 1/2.0, 0/1.8}{
  \draw[draw=colU, fill=colU!8, thick]
    ({6.0+0.22*\i}, {-0.5-0.22*\i}) rectangle ({7.0+0.22*\i}, {-0.5-0.22*\i-\h});
}
\node[font=\footnotesize] at (6.5, -1.5) {$\bm{U}_k$};
 
\draw[decorate, decoration={brace, amplitude=5pt, raise=3pt, mirror}]
  (6.0, -0.5) -- (6.0, -2.3)
  node[midway, left=8pt, font=\footnotesize] {$I_k$};
\draw[decorate, decoration={brace, amplitude=5pt, raise=3pt}]
  (6.0, -0.5) -- (7.0, -0.5)
  node[midway, above=8pt, font=\footnotesize] {$R$};

\foreach \i in {4,3,2,1,0}{
  \draw[draw=colS, fill=colS!8, thick]
    ({8.0+0.18*\i}, {-0.5-0.18*\i}) rectangle ({9.0+0.18*\i}, {-1.7-0.18*\i});
}
\foreach \d in {1.2, 0.6, 0, -0.6, -1.2}{
   \node[fill=colS!50, inner sep=2.2pt, rectangle] 
    at ({8.5-\d*0.28}, {-1.1+\d*0.38}) {};
}

\node[font=\footnotesize] at (8.55, -1.1) {$\bm{S}_k$};
 
\draw[decorate, decoration={brace, amplitude=5pt, raise=3pt}]
  (8.0, -0.5) -- (9.0, -0.5)
  node[midway, above=8pt, font=\footnotesize] {$R$};

\draw[draw=colV, fill=colV!8, thick]
  (10.7, -0.5) rectangle (11.7, -4.0);
\node[font=\footnotesize] at (11.2, -2.25) {$\bm{V}$};
 
\draw[decorate, decoration={brace, amplitude=5pt, raise=3pt}]
  (11.7, -0.5) -- (11.7, -4.0)
  node[midway, right=8pt, font=\footnotesize] {$J$};
\draw[decorate, decoration={brace, amplitude=5pt, raise=3pt}]
  (10.7, -0.5) -- (11.7, -0.5)
  node[midway, above=8pt, font=\footnotesize] {$R$};

\draw[draw=colW, fill=colW!8, thick]
  (8.2, -5.0) rectangle (9.2, -9.0);
\node[font=\footnotesize] at (8.75, -6.9)
  {
   {\footnotesize$\bm{W}$}};

\draw[decorate, decoration={brace, amplitude=5pt, raise=3pt, mirror}]
  (8.2, -5.0) -- (8.2, -9.0)
  node[midway, left=8pt, font=\footnotesize] {$K$};
 
\draw[->, colS, thick, dashed]
  (8.6, -2.4) -- (8.6, -4.9)
  node[midway, right=0pt, font=\tiny, colS] {$\boldsymbol{w}_k = \mathrm{diag}(\bm{S}_k)$};

\draw[draw=colCox, fill=colCox!6, thick, rounded corners=5pt]
  (12.8, -2.0) rectangle (15.4, -3.2);
\node[font=\footnotesize] at (14.1, -2.6)
  {$\mathcal{L}(\boldsymbol{\beta},\boldsymbol{\gamma};\,\mathcal{H} \cup \mathcal{C})$};

\draw[draw=colRank, fill=colRank!5, thick, rounded corners=5pt]
  (12.8, -4.0) rectangle (15.4, -8.1);
\node[font=\footnotesize\bfseries, colRank] at (14.1, -4.3) {Risk ranking};

\foreach \bh/\by/\col in {
  1.6/-5.0/colH,
  1.2/-5.7/colC,
  0.8/-6.4/colC,
  0.6/-7.1/colH,
  0.4/-7.8/colH}{
  \draw[fill=\col!60, draw=\col!80, thin]
    (13.0, \by) rectangle ({13.0+\bh}, {\by+0.5});
}

\draw[->, colCox, thick]
  (14.1, -3.2) -- (14.1, -4.0)
  node[midway, right=4pt, font=\tiny] {$\hat{\boldsymbol{\beta}}, \hat{\boldsymbol{\gamma}}$};
 
\draw[->, colC, thick, dashed]
  (9.4, -6.2) -- (12.8, -6.2)
  node[midway, above, font=\small, colC] {$\mathcal{H} \cup \mathcal{C}$};

\node[font=\footnotesize\bfseries] at (1.5, -9.0)
  {Input};
\node[font=\footnotesize\bfseries] at (9.0, -9.0)
  {};

\end{tikzpicture}

\caption{Overview of PAR2COX: historical and current cohort tensors are decomposed using our joint optimization algorithm to extract patient-specific latent factors, which serve as additional covariates in the Cox log-likelihood.}
\label{fig:par2cox}
\end{figure}

\setlength{\baselineskip}{\normalbaselineskip}

\subsection{Selection of Tuning Parameters}\label{sec:tuning}
PAR2COX requires selecting the decomposition rank $R$, the regularization parameters $\alpha$ and $\eta$, and the parameter $\lambda$ controlling the contribution of the Cox log-likelihood term. The rank $R$ may be selected either based on prior domain knowledge (e.g., expected biological or clinical structure) or empirically through validation-based model selection. In the absence of prior information, we recommend selecting $R$ using the same data-driven tuning strategy adopted for the remaining parameters.
To select the tuning parameters, we use the concordance index (C-index) \citep{harrell1982} evaluated on held-out data (see Section \ref{sec:sim} for details on C-index). Based on our experiments, we recommend a two-stage strategy for parameter tuning. First, an initial exploration of the parameter space is performed using Latin hypercube sampling, which allows efficient coverage of the multidimensional search space. Subsequently, a refined grid search is conducted in regions associated with promising performance. The parameter combination achieving the highest C-index in the refined search is selected as the final set of tuning parameters.

\subsection{Computational Complexity}

In this section, we study the computational complexity of the proposed method. Let $E$ denote the number of events and, for the $i-$th event, let $b_i$
denote the size of its risk set (i.e., the number of patients under
observation at the corresponding event time). We further define $B = \sum_{i=1}^{E} b_i$, the total risk-set membership across all
events. Finally, let $T_{out}$ denote the number of outer iterations, $T_{in}$ the number of inner iterations
used to update $\{\bm{H},\bm{V},\bm{W},\boldsymbol{\beta},\boldsymbol{\gamma}\}$ for fixed $\{\bm{Q}_k\}$, $T_{fs}$ the
number of iterations required by the nonlinear solver used to update each
row of $\bm{W}$, and $T_{qn}$ the number of quasi-Newton iterations used to
update $\boldsymbol{\beta}$ and $\boldsymbol{\gamma}$.

Updating $\bm{Q}_k$ for a fixed $k$ requires computing $\bm{X}_k\bm{V}\bm{S}_k\bm{H}^\top$, at a cost of $O(I_kJR+I_kR^2)$, while the cost of the subsequent truncated SVD step is $O(I_kR^2)$. Since $J > R$, the cost of updating $\bm{Q}_k$ reduces to $O(I_kJR)$. Repeating this procedure for all $K$ patients yields a total
cost of $O(I_{tot}JR)$ per outer iteration, with $I_{tot} = \sum_{k=1}^K I_k$.

With $\{\bm{Q}_k\}$ fixed, the cost of updating $\bm{H}$ is $O(JKR^2)$ per inner iteration. In addition, the costs of updating $\overline{\bm{H}}$ and $\bm{\Gamma}_H$ when nonnegativity constraint is added are both $O(R^2)$. Hence, the total cost associated with the updates in (\ref{eq:updateH}) is $O(JKR^2 + R^2)$, which reduces to $O(JKR^2)$. Similarly, the cost of the $\bm{V}$ update is $O(JKR^2)$, while the costs of $\overline{\bm{V}}$ and $\bm{\Gamma}_V$ are both $O(JR)$. Therefore, the total cost associated with the updates in (\ref{eq:updateV}) is $O(JKR^2 + JR)$, which reduces to $O(JKR^2)$.

The cost of updating $\bm{W}$ at each inner iteration is the sum of two contributions: a precomputation cost, incurred once per iteration, and a per-subject cost, incurred for each patient. The precomputation includes the computation of  $(\bm{V} \odot \bm{H})$ and its Gram matrix, with costs $O(JR^2)$ and $O(JR^2 + R^3)$, which reduces to $O(JR^2)$, respectively. In addition, we precompute the linear predictors for all patients, sort the patients by event time, and compute all risk set denominators with costs 
$O(K(R+P))$, $O(K\log K)$, and  $O(K)$. Therefore, the total precomputation cost is $O(JR^2 + K(R+P) + K\log K)$. 
The cost of updating the $i-$th row of $\bm{W}$ is 
$O(JR^2 + T_{fs}( R^3 + Rb_i))$. Hence, updating all $K$ rows of $\bm{W}$ costs $O(KJR^2 + T_{fs}( KR^3 + RB))$. 
In addition, the costs of updating $\overline{\bm{W}}$ and $\bm{\Gamma}_W$ are both $O(KR)$. Lastly, the cost of updating $\boldsymbol{\theta} = (\boldsymbol{\beta}^\top, \boldsymbol{\gamma}^\top)$ is $O(K \log K + K(R + P) + T_{qn} B (R + P)^2)$. 

Therefore, the total computational complexity of PAR2COX is 
$O(T_{out}[I_{tot}JR + T_{in}(KJR^2 + T_{fs}(KR^3 + RB) + T_{qn} B (R+P)^2+ K\log K)])$. The complexity is linear in the cohort size $K$, the total number of visits $I_{tot}$, and the feature count $J$, and only low-order polynomial in the small quantities $R$ and $P$. 

\section{Simulation Study}\label{sec:sim}
In this section, we assess the risk prediction performance of PAR2COX through an extensive simulation study. PAR2COX is compared with four baseline methods: PCA \citep{jolliffe}, CP-WOPT \citep{cpwopt}, DTW-CP \citep{dtwcp}, and COPA \citep{copa}.
In particular, to apply PCA, each slice $\bm{X}_k \in \mathbb{R}^{I_k \times J}$ is first reduced to a fixed-length vector $\boldsymbol{x}_k \in \mathbb{R}^{J}$ by averaging across rows, whereas the remaining baselines operate directly on higher-order tensors.  

In all simulation scenarios, the longitudinal data for patient $k$ are given by $\mathbf{X}_k \in \mathbb{R}^{I_k \times J}$, where $I_k \sim U\{I_{\min}, I_{\max}\}$ is the number of visits and $J$ is the number of features. The data are generated according to the PARAFAC2 model
\begin{equation*}
    \mathbf{X}_k = \mathbf{P}_k \mathbf{H} \, \bm{S}_k \, \mathbf{V}^\top
\end{equation*}
where $\mathbf{P}_k \in \mathbb{R}^{I_k \times R}$ is a random orthonormal matrix, $\mathbf{H} \in \mathbb{R}^{R \times R}$ is a fixed orthonormal matrix shared across patients, $\bm{S}_k = \mathrm{diag}(\boldsymbol{w}_k) \in \mathbb{R}^{R \times R}$ with the patient-specific latent factor scores $\boldsymbol{w}_k \in \mathbb{R}^R$, and $\mathbf{V} \in \mathbb{R}^{J \times R}$ is a fixed factor matrix with entries drawn from $U(0,1)$. The latent scores are generated according to a two-group structure that reflects a population of high- and low-risk patients. A proportion $\pi_h$ of patients is assigned to the high-risk group, for which $w_{kr} \sim U(n_0, n_0 + 1)$ with $n_0 > 0$, while the remaining patients are in the low-risk group with $w_{kr} \sim U(0, 1)$, for $r = 1, \dots R$ and $k = 1, \dots, K$. The offset $n_0$ controls the separation between the two groups, with smaller values inducing greater overlap and hence a more challenging risk-stratification task. Group membership is assigned at random, and the ordering of subjects is subsequently permuted to avoid introducing any systematic structure into the data. The coefficient vector $\boldsymbol{\beta} \in \mathbb{R}^R$ is generated as $\beta_r \sim U(\beta_0, \beta_0 + 1)$, for $r = 1, \dots, R$, where the offset parameter $\beta_0$ governs the overall strength of the association between the latent scores and the hazard.

Survival times are generated according to the procedure described by \citet{bender2005generating} for the Cox proportional hazards model with exponential baseline hazard. For patient $k$, the survival time is given by
\begin{equation*}
    D_k = \frac{-\log U_k}{\lambda_0 \, \exp ( \boldsymbol{w}_k^\top \boldsymbol{\beta}  +  \boldsymbol{z}_k^\top \boldsymbol{\gamma})} 
\end{equation*}
where $U_k \sim U(0,1)$ and $\lambda_0 > 0$ is the baseline hazard rate. 
For simplicity, we exclude time-invariant covariates from the scenarios presented below, (i.e., $\boldsymbol{z}_k^{\top} \boldsymbol{\gamma} = 0$), and $\boldsymbol{\gamma}$ is therefore not generated.
The $K$ patients are then randomly partitioned into a historical cohort $\mathcal{H}$ and a current cohort $\mathcal{C}$, with $|\mathcal{H}| = \lfloor K(1 - \pi_\mathcal{C}) \rfloor$ and $|\mathcal{C}| = K - |\mathcal{H}|$, where $\pi_\mathcal{C}$ is the proportion of patients currently under observation.
% censoring 
Non-informative random censoring is applied to the historical cohort, with censoring times drawn independently from an exponential distribution whose rate is chosen to achieve a pre-specified censoring rate (CR). In the current cohort, all patients are treated as censored at the time of analysis; their true survival times are withheld during model fitting and used only for evaluation, reflecting that their outcomes are not yet observed in practice.

To assess risk prediction performance, we adopt the C-index \citep{harrell1982}, defined as 
\begin{equation}\label{eq:cindex}
   \widehat{C} = \frac{\sum_{k \in O} \sum_{j = k+1}^K I(t_k < t_j)\, I(M_k > M_j)}{\sum_{k \in O} \sum_{j = k+1}^K I(t_k < t_j)}
\end{equation}
under the assumption of no ties. In \eqref{eq:cindex}, $M_k$ and $M_j$ denote the risk scores of patients $k$ and $j$, respectively, and $I(\cdot)$ is the indicator function. The C-index measures the proportion of concordant pairs among all comparable pairs. A pair of patients is comparable if it can be determined which patient experienced the event first. A comparable pair is also concordant if the ordering of the predicted risk scores agrees with the observed ordering of the event times. The C-index is a standard performance metric in survival analysis that quantifies the agreement between predicted risks and observed event times. A value of $\widehat{C} = 1$ indicates perfect concordance, while $\widehat{C} = 0.5$ corresponds to random risk assignment. 
In the current setting, the estimated risk score for patient $k$ is given by $M_k = \boldsymbol{w}_k^\top {\boldsymbol{\beta}} + \boldsymbol{z}_k^\top {\boldsymbol{\gamma}}$.  

In the simulations, we set $K = 300$ patients, $J = 20$ clinical features, $R = 3$ latent factors, and minimum and maximum number of visits equal to $I_{\min} = 40$ and $I_{\max} = 60$, respectively. Furthermore, the baseline hazard rate is set to $\lambda_0 = 0.05$, the proportion of high-risk patients is fixed at $\pi_h = 0.2$ and the proportion of patients belonging to the current cohort is set to $\pi_\mathcal{C} = 0.2$, yielding $|\mathcal{C}| = 60$. 

We use both COPA and PAR2COX without imposing constraints, which reduces COPA to the standard PARAFAC2 decomposition. PAR2COX is fitted with a maximum of $200$ outer iterations, $5$ ADMM inner iterations, and tuning parameters (i.e., $\alpha$, $\eta$, $\lambda$) selected via Latin hypercube sampling followed by grid search on held-out data generated from the same mechanism (see Section \ref{sec:tuning}). In addition to the four baseline methods, the performance of the Oracle is reported (i.e., the C-index computed from the true $\boldsymbol{\beta}$ and $\boldsymbol{w}_k$, for $k = 1, \dots, K$). 

The performance of PAR2COX and competing methods is evaluated across different values of $\beta_0$ and $n_0$. In particular, we consider $\beta_0 = 1.0$ and $\beta_0 = 2.0$, and $n_0 = 0.5$ and $n_0 = 0.3$. As an example, $n_0 = 0.5$ results in $w_{kr} \sim U(0, 1)$ for low-risk patients and $w_{kr} \sim U(0.5, 1.5)$ for high-risk patients. Notably, these values of $n_0$ yield substantial overlap in the distributions of the latent factor scores across risk groups.
Results are reported for $\mathcal{H} \, \cup \, \mathcal{C}$ and $\mathcal{C}$ across censoring rates $\{0\%, 10\%, 20\%, 30\%\}$.
The first setting reflects a clinically relevant scenario in which the objective is to rank patients across both historical and current cohorts. For example, such rankings may be used to assess whether current patients are at higher or lower risk relative to historical patients, thereby supporting risk stratification and clinical decision-making. 
A ranking restricted to patients in $\mathcal{C}$ can be obtained from the overall ranking. However, in the second setting, performance is evaluated using the C-index computed on $\mathcal{C}$, with the true survival times of patients in $\mathcal{C}$ used for evaluation. Evaluating the ability of the methods to correctly rank patients within the current cohort is relevant when the objective is to prioritize patients for clinical interventions, treatment or transplant allocation. 

\begin{table}[t]
\centering
\caption{C-index mean (standard error) across 1000 repetitions, $\beta_0 = 1.0$, $n_0 = 0.5$. Best results (excluding Oracle) in bold.}
\label{tab:results_beta1}

\resizebox{\textwidth}{!}{%
\begin{tabular}{lccccc|c}
\toprule
\textbf{$\mathcal{H} \, \cup \, \mathcal{C}$} & & &\\

\multicolumn{7}{c}{\textbf{Methods}}\\
\midrule
\textbf{CR} & PCA & CP-WOPT & DTW-CP & COPA & PAR2COX & Oracle \\
\midrule
$0\%$  & 0.5336 (0.0005) & 0.5336 (0.0006) & 0.7065 (0.0015) & 0.7388 (0.0008) & \textbf{0.7508 (0.0007)} & 0.7523 (0.0008) \\
$10\%$ & 0.5345 (0.0006) & 0.5351 (0.0006) & 0.7118 (0.0015) & 0.7429 (0.0008) & \textbf{0.7550 (0.0008)} & 0.7563 (0.0008) \\
$20\%$ & 0.5371 (0.0006) & 0.5372 (0.0007) & 0.7178 (0.0015) & 0.7479 (0.0009) & \textbf{0.7599 (0.0008)} & 0.7611 (0.0008) \\
$30\%$ & 0.5398 (0.0007) & 0.5397 (0.0007) & 0.7250 (0.0015) & 0.7535 (0.0009) & \textbf{0.7657 (0.0008)} & 0.7667 (0.0008) \\
\midrule
$\mathcal{C}$ & & &\\
\multicolumn{7}{c}{\textbf{Methods}}\\
\midrule
\textbf{CR} & PCA & CP-WOPT & DTW-CP & COPA & PAR2COX & Oracle \\
\midrule
$0\%$  & 0.5062 (0.0014) & 0.5075 (0.0015) & 0.6982 (0.0018) & 0.7198 (0.0013) & \textbf{0.7433 (0.0012)} & 0.7453 (0.0012) \\
$10\%$ &  0.5052 (0.0014) & 0.5074 (0.0015) & 0.6982 (0.0018) & 0.7200 (0.0013) & \textbf{0.7432 (0.0012)} & 0.7453 (0.0012) \\
$20\%$ & 0.5064 (0.0014) & 0.5089 (0.0015) & 0.6981 (0.0018) & 0.7200 (0.0013) & \textbf{0.7431 (0.0012)} & 0.7453 (0.0012) \\
$30\%$ & 0.5054 (0.0014) & 0.5077 (0.0015) & 0.6976 (0.0018) & 0.7200 (0.0013) & \textbf{0.7429 (0.0012)} & 0.7453 (0.0012) \\
\bottomrule
\end{tabular}%
}

\end{table}

\begin{table}[t]
\centering
\caption{C-index mean (standard error) across 1000 repetitions, $\beta_0 = 2.0$, $n_0 = 0.5$. Best results (excluding Oracle) in bold.}
\label{tab:results_beta2}
\resizebox{\textwidth}{!}{%
\begin{tabular}{lccccc|c}
\toprule
\textbf{$\mathcal{H} \, \cup \, \mathcal{C}$} & & &\\
\multicolumn{7}{c}{\textbf{Methods}}\\
\midrule
\textbf{CR} & PCA & CP-WOPT & DTW-CP & COPA & PAR2COX & Oracle \\
\midrule
$0\%$ & 0.5348 (0.0006) & 0.5353 (0.0006) & 0.7603 (0.0018) & 0.8054 (0.0007) & \textbf{0.8235 (0.0005)} & 0.8263 (0.0005) \\
$10\%$ & 0.5361 (0.0006) & 0.5368 (0.0007) & 0.7669 (0.0018) & 0.8098 (0.0007) & \textbf{0.8282 (0.0005)} & 0.8308 (0.0005) \\
$20\%$ &   0.5376 (0.0006) & 0.5389 (0.0007) & 0.7751 (0.0018) & 0.8154 (0.0007) & \textbf{0.8340 (0.0005)} & 0.8364 (0.0005) \\
$30\%$ & 0.5409 (0.0007) & 0.5414 (0.0007) & 0.7849 (0.0018) & 0.8225 (0.0007) & \textbf{0.8412 (0.0005)} & 0.8434 (0.0005) \\
\midrule
$\mathcal{C}$ & & &\\
\multicolumn{7}{c}{\textbf{Methods}}\\
\midrule
\textbf{CR} & PCA & CP-WOPT & DTW-CP & COPA & PAR2COX & Oracle \\
\midrule
$0\%$  & 0.5061 (0.0015) & 0.5071 (0.0015) & 0.7519 (0.0021) & 0.7830 (0.0012) & \textbf{0.8153 (0.0009)} & 0.8184 (0.0009) \\
$10\%$ & 0.5057 (0.0015) & 0.5073 (0.0015) & 0.7517 (0.0021) & 0.7830 (0.0012) & \textbf{0.8152 (0.0009)} & 0.8184 (0.0009) \\
$20\%$ & 0.5053 (0.0014) & 0.5072 (0.0015) & 0.7516 (0.0021) & 0.7830 (0.0012) & \textbf{0.8152 (0.0009)} & 0.8184 (0.0009) \\
$30\%$ &   0.5062 (0.0015) & 0.5073 (0.0015) & 0.7510 (0.0021) & 0.7831 (0.0012) & \textbf{0.8152 (0.0009)} & 0.8184 (0.0009) \\
\bottomrule
\end{tabular}%
}
\end{table}

\begin{table}[t]
\centering
\caption{C-index mean (standard error) across 1000 repetitions, $\beta_0 = 1.0$, $n_0 = 0.3$. Best results (excluding Oracle) in bold.}
\label{tab:results_beta1_n03}

\resizebox{\textwidth}{!}{%
\begin{tabular}{lccccc|c}
\toprule
\textbf{$\mathcal{H} \, \cup \, \mathcal{C}$} & & &\\

\multicolumn{7}{c}{\textbf{Methods}}\\
\midrule
\textbf{CR} & PCA & CP-WOPT & DTW-CP & COPA & PAR2COX & Oracle \\
\midrule
$0\%$  &  0.5326 (0.0005) & 0.5322 (0.0006) & 0.6780 (0.0013) & 0.7080 (0.0008) & \textbf{0.7201 (0.0008)} & 0.7214 (0.0008) \\
$10\%$ & 0.5340 (0.0006) & 0.5336 (0.0006) & 0.6814 (0.0014) & 0.7104 (0.0009) & \textbf{0.7225 (0.0008)} & 0.7237 (0.0008) \\
$20\%$ & 0.5353 (0.0006) & 0.5360 (0.0006) & 0.6855 (0.0014) & 0.7133 (0.0009) & \textbf{0.7257 (0.0008)} & 0.7266 (0.0009) \\
$30\%$ & 0.5384 (0.0006) & 0.5381 (0.0007) & 0.6897 (0.0014) & 0.7168 (0.0009) & \textbf{0.7289 (0.0009)} & 0.7297 (0.0009) \\
\midrule
$\mathcal{C}$ & & &\\
\multicolumn{7}{c}{\textbf{Methods}}\\
\midrule
\textbf{CR} & PCA & CP-WOPT & DTW-CP & COPA & PAR2COX & Oracle \\
\midrule
$0\%$  & 0.5072 (0.0014) & 0.5089 (0.0015) & 0.6706 (0.0017) & 0.6894 (0.0013) & \textbf{0.7131 (0.0013)} & 0.7153 (0.0013) \\
$10\%$ & 0.5069 (0.0014) & 0.5080 (0.0014) & 0.6708 (0.0017) & 0.6894 (0.0013) & \textbf{0.7131 (0.0013)} & 0.7153 (0.0013) \\
$20\%$ & 0.5078 (0.0014) & 0.5074 (0.0014) & 0.6709 (0.0017) & 0.6895 (0.0013) & \textbf{0.7130 (0.0013)} & 0.7153 (0.0013) \\
$30\%$ & 0.5081 (0.0013) & 0.5081 (0.0014) & 0.6708 (0.0017) & 0.6893 (0.0013) & \textbf{0.7128 (0.0013)} & 0.7153 (0.0013) \\
\bottomrule
\end{tabular}%
}

\end{table}

\begin{figure}[htb!]
    \centering
    \begin{subfigure}[b]{0.49\textwidth}
        \centering
        \includegraphics[width=\textwidth]{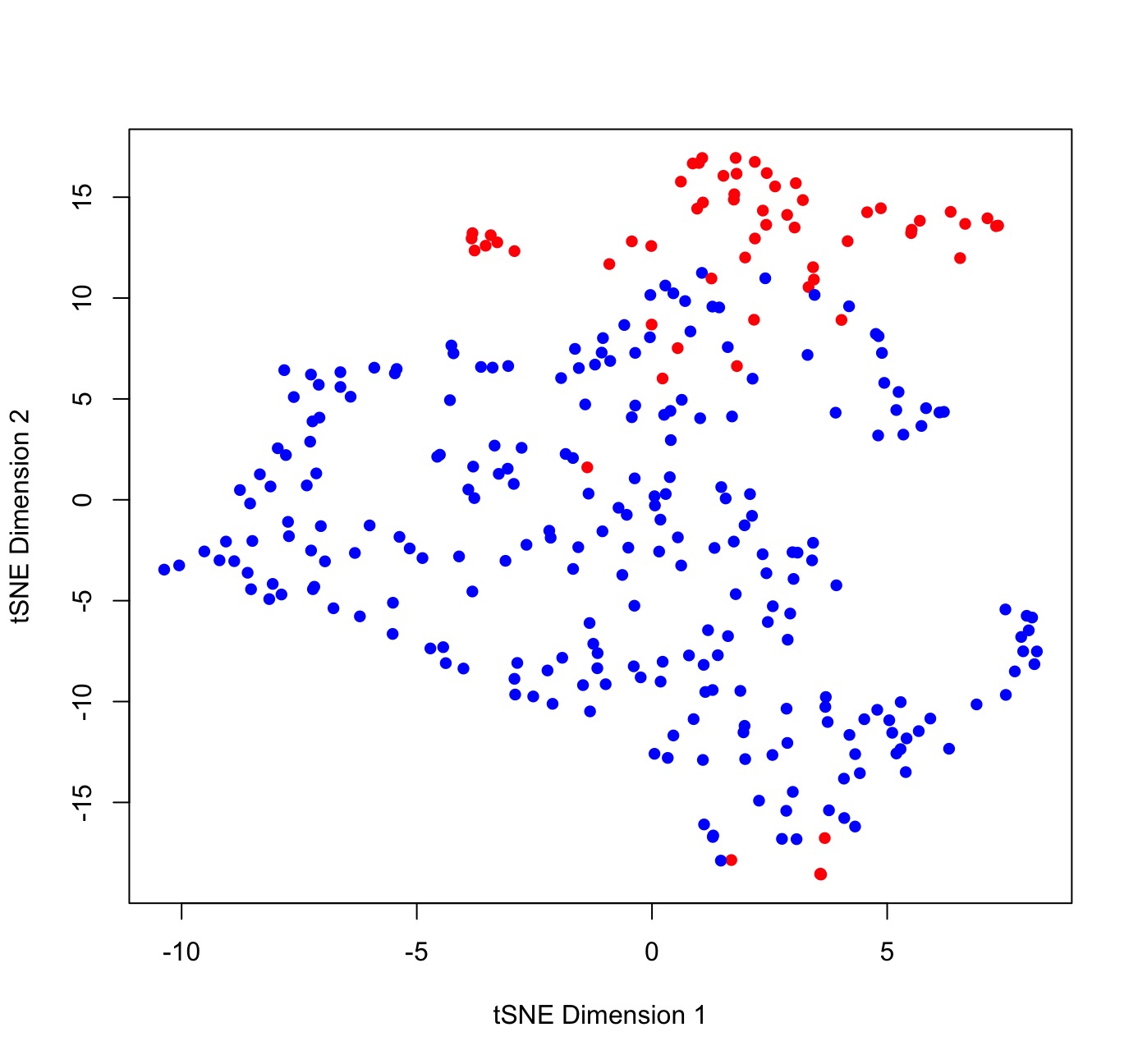}
        \caption{COPA, $n_0 = 0.5$}
        \label{fig:vis-copa05}
    \end{subfigure}
    \hfill
    \begin{subfigure}[b]{0.49\textwidth}
        \centering
        \includegraphics[width=\textwidth]{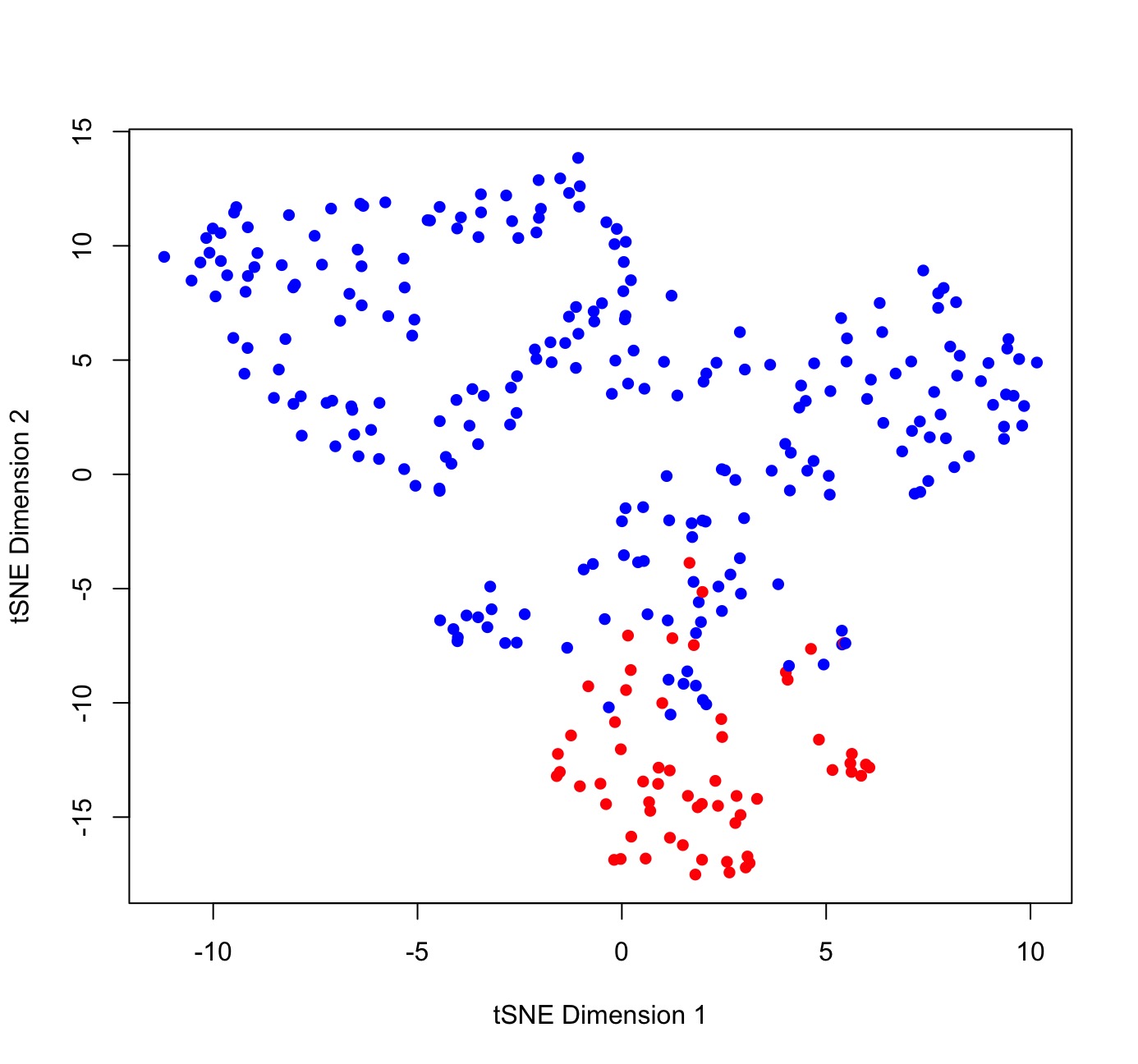}
        \caption{PAR2COX, $n_0 = 0.5$}
        \label{fig:vis-par2cox05}
    \end{subfigure}

    \vspace{0.5em}

    \begin{subfigure}[b]{0.49\textwidth}
        \centering
        \includegraphics[width=\textwidth]{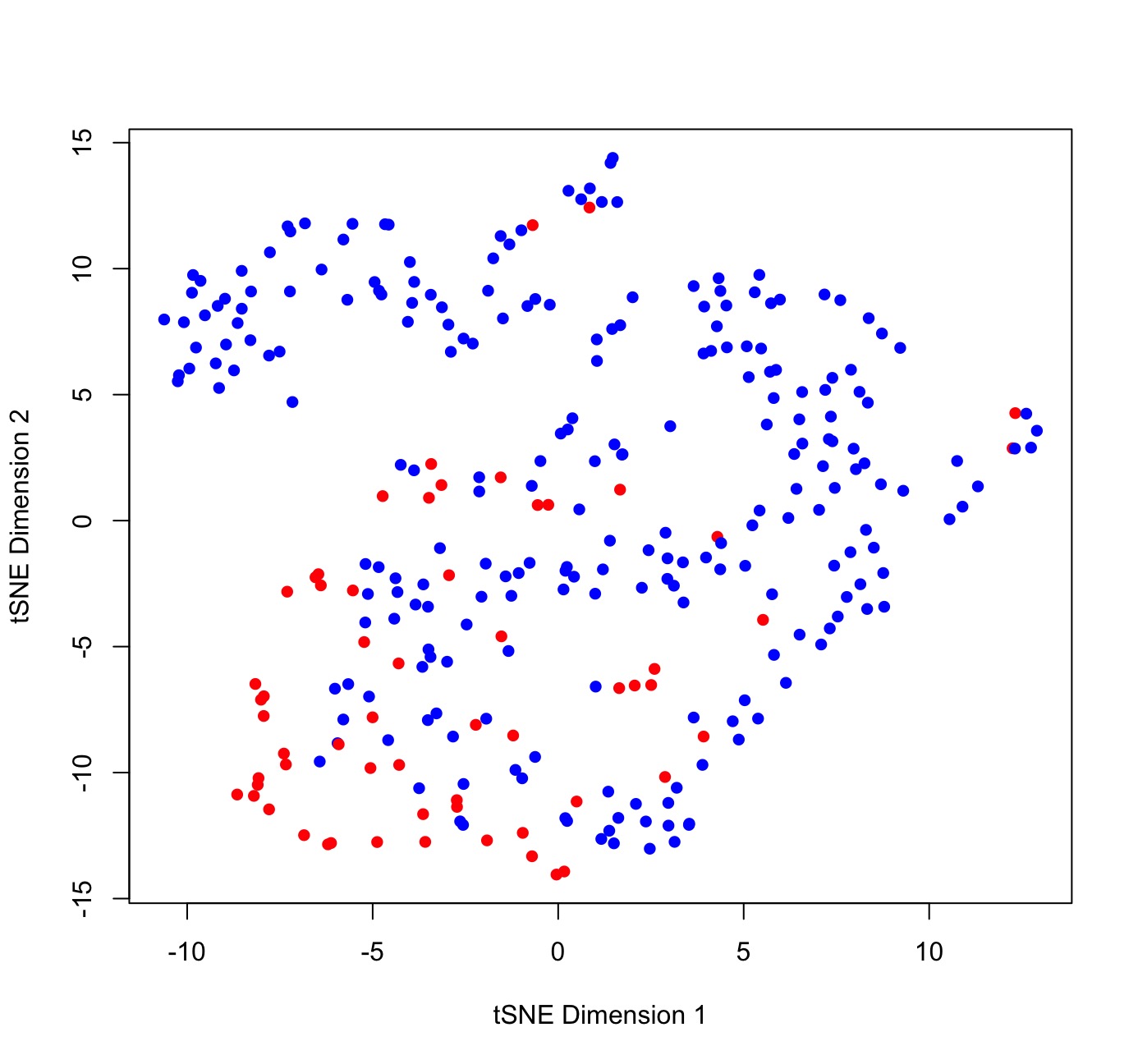}
        \caption{COPA, $n_0 = 0.3$}
        \label{fig:vis-copa09}
    \end{subfigure}
    \hfill
    \begin{subfigure}[b]{0.49\textwidth}
        \centering
        \includegraphics[width=\textwidth]{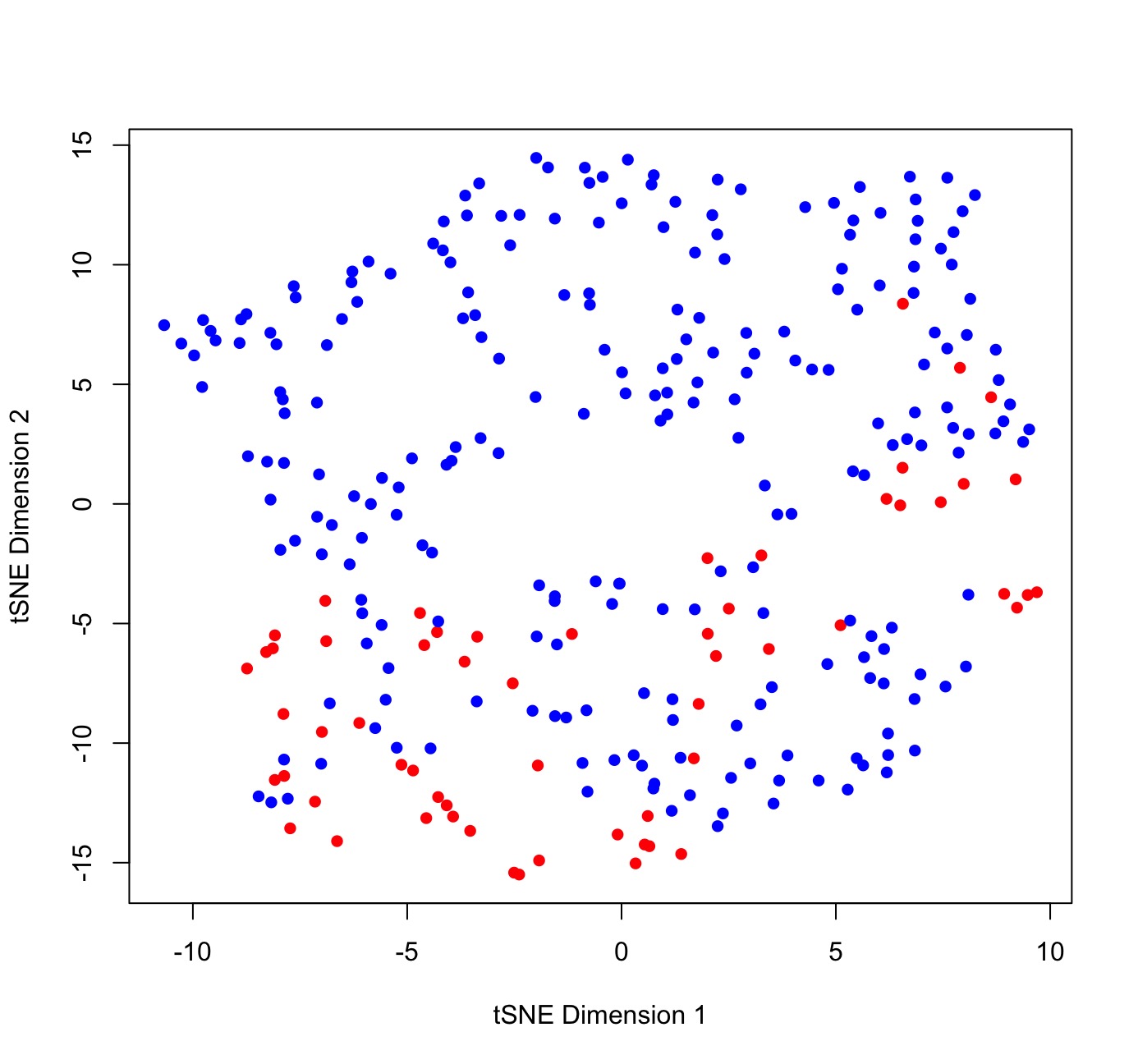}
        \caption{PAR2COX, $n_0 = 0.3$}
        \label{fig:vis-par2cox09}
    \end{subfigure}

    \caption{tSNE mapping for COPA and PAR2COX (high-risk subjects in red, low-risk subjects in blue).}
    \label{fig:vis-all}
\end{figure}

Tables \ref{tab:results_beta1} and \ref{tab:results_beta2} present the results for $n_0 = 0.5$ with $\beta_0 = 1.0$ and $\beta_0 = 2.0$, respectively. 
The performance of all methods remains relatively stable across CR levels, indicating that the censoring level has a limited impact on risk stratification.
In both scenarios and across all censoring rates, PAR2COX achieves the highest C-index. Notably, its performance closely approaches the Oracle values, and consistently outperforms the best competing method (i.e., COPA). 
In Tables \ref{tab:results_beta1_n03} and \ref{tab:results_beta2_n03}, results are reported for the more challenging scenarios with $n_0 = 0.3$, and $\beta_0 = 1.0$ and $\beta_0 = 2.0$, respectively. As expected, reducing  $n_0$ increases the overlap between the high- and low-risk groups, making the risk stratification task more challenging. Although the performance of all methods deteriorates, PAR2COX consistently achieves the highest C-index across all experimental settings, demonstrating its robustness even when the latent group structure is less distinct. Overall, PAR2COX consistently outperforms all competing methods, indicating that incorporating survival information during representation learning improves risk stratification performance. 
In addition, we report the tSNE visualizations of the extracted patient-specific scores in Figure \ref{fig:vis-all}. Specifically, the data are generated according to the second scenario (i.e., $\beta_0 = 2.0$), and the visualization is reported for PAR2COX and COPA. 
Each point corresponds to a patient, while the red and blue colors denote membership in the true high- and low-risk groups, respectively. Although our choices for $n_0$ lead to substantial overlapping between the two groups, PAR2COX achieves a clearer separation of the patient clusters than COPA, highlighting its superior ability to recover the underlying clinical condition even under challenging settings.

\begin{table}[t]
\centering
\caption{C-index mean (standard error) across 1000 repetitions, $\beta_0 = 2.0$, $n_0 = 0.3$. Best results (excluding Oracle) in bold.}
\label{tab:results_beta2_n03}

\resizebox{\textwidth}{!}{%
\begin{tabular}{lccccc|c}
\toprule
\textbf{$\mathcal{H} \, \cup \, \mathcal{C}$} & & &\\

\multicolumn{7}{c}{\textbf{Methods}}\\
\midrule
\textbf{CR} & PCA & CP-WOPT & DTW-CP & COPA & PAR2COX & Oracle \\
\midrule
$0\%$  & 0.5336 (0.0005) & 0.5340 (0.0006) & 0.7353 (0.0017) & 0.7776 (0.0007) & \textbf{0.7963 (0.0006)} & 0.7995 (0.0006) \\
$10\%$ & 0.5349 (0.0006) & 0.5353 (0.0006) & 0.7405 (0.0017) & 0.7808 (0.0007) & \textbf{0.7997 (0.0006)} & 0.8027 (0.0006) \\
$20\%$ & 0.5369 (0.0006) & 0.5374 (0.0006) & 0.7467 (0.0017) & 0.7845 (0.0008) & \textbf{0.8037 (0.0006)} & 0.8066 (0.0006) \\
$30\%$ & 0.5392 (0.0006) & 0.5401 (0.0007) & 0.7541 (0.0017) & 0.7896 (0.0008) & \textbf{0.8089 (0.0006)} & 0.8115 (0.0006) \\
\midrule
$\mathcal{C}$ & & &\\
\multicolumn{7}{c}{\textbf{Methods}}\\
\midrule
\textbf{CR} & PCA & CP-WOPT & DTW-CP & COPA & PAR2COX & Oracle \\
\midrule
$0\%$  & 0.5070 (0.0014) & 0.5085 (0.0015) & 0.7274 (0.0020) & 0.7547 (0.0012) & \textbf{0.7883 (0.0010)} & 0.7919 (0.0010) \\
$10\%$ & 0.5068 (0.0014) & 0.5079 (0.0015) & 0.7276 (0.0019) & 0.7546 (0.0012) & \textbf{0.7881 (0.0010)} & 0.7919 (0.0010) \\
$20\%$ & 0.5063 (0.0014) & 0.5090 (0.0015) & 0.7277 (0.0019) & 0.7547 (0.0012) & \textbf{0.7879 (0.0010)} & 0.7919 (0.0010) \\
$30\%$ & 0.5063 (0.0014) & 0.5075 (0.0015) & 0.7278 (0.0019) & 0.7546 (0.0012) & \textbf{0.7881 (0.0010)} & 0.7919 (0.0010) \\
\bottomrule
\end{tabular}%
}

\end{table}

\clearpage 

\clearpage 

\section{Case Study}\label{sec:casestudy}
Sepsis remains one of the leading causes of morbidity and mortality worldwide, imposing a substantial burden on healthcare systems and ICU.
In 2017, sepsis accounted for an estimated 48.9 million incident cases and 11.0 million deaths worldwide, representing approximately the $20\%$ of all global deaths \citep{Rudd2020Sepsis}. As sepsis is characterized by progressive organ dysfunction resulting from a dysregulated response to infection \citep{singer2016third}, patients in ICU are routinely monitored through repeated clinical assessments during their stay. 
The resulting longitudinal data often capture a highly heterogeneous and rapidly evolving clinical course, highlighting the need for statistical methods capable of accurately modeling disease progression, stratifying patient risk, and supporting clinical decision-making. 

\begin{figure}[t]
    \centering
    \includegraphics[width=0.5\linewidth]{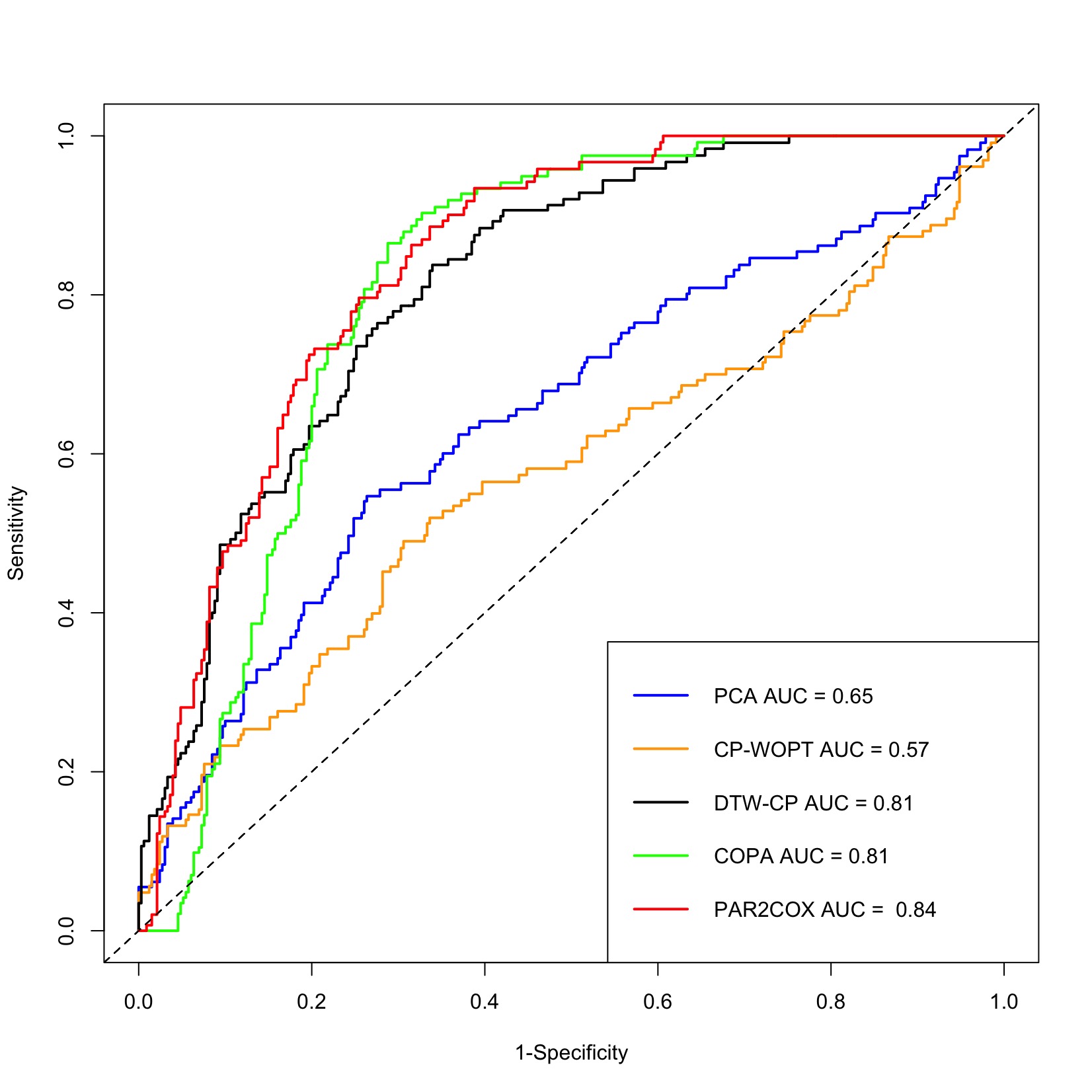}
    \caption{Time-dependent ROC curves for all methods at the median survival time, with corresponding AUC values.}
    \label{fig:timeROC}
\end{figure}

These challenges motivate the application of our proposed approach to a sepsis cohort extracted from the MIMIC-IV database \citep{Johnson2023paper, Johnson2024MIMICIV}. The cohort consists of $K = 700$ patients and $J = 59$ variables, which include $(i)$ vital signs such as temperature, heart and respiratory rate, $(ii)$  laboratory measurements such as lactate, glucose and hemoglobin, and $(iii)$ comorbidities such as hypertension, diabetes, chronic lung disease, cardiovascular disease, kidney and heart failures. In addition, the dataset includes the Sequential Organ Failure Assessment (SOFA) score, a composite measure of the degree of dysfunction across six organ systems (i.e., respiratory, cardiovascular, hepatic, coagulation, renal, and neurological) \citep{Vincent1996SOFA, singer2016third}. We split the patients into historical and current cohorts. The historical cohort includes 560 patients, 284 of which are censored, while the current cohort includes 140 censored patients. To ensure a fair comparison, we apply PAR2COX without imposing additional constraints on the factor matrices and evaluate all competing methods listed in Section \ref{sec:sim} under the same experimental setting. The selected tuning parameters for PAR2COX are $R = 4$, $\alpha = 0.50$, $\eta = 0.50$ and $\lambda = 0.02$. Table \ref{tab:par2cox-casestudy} reports the C-index values computed on the historical and the combined cohorts for all methods. 
 Here, the C-index cannot be evaluated for $\mathcal{C}$ alone, since in a real-world setting the true event times of patients who are under observation are unavailable.
 PAR2COX achieves the highest C-index for both $\mathcal{H}$ and $\mathcal{H} \cup \mathcal{C}$, demonstrating superior risk prediction performance in a real-world clinical setting. 
Furthermore, Figure \ref{fig:timeROC} presents the time-dependent ROC curves \citep{heagerty2000time} for all methods and their corresponding area under the curve (AUC) values. PAR2COX attains the highest AUC of $0.84$, followed by COPA and DTW-CP, both of which achieve AUC$ = 0.81$. Overall, these results suggest that PAR2COX may offer improved risk stratification in real-world clinical settings. 
\begin{table}[t]
\centering
\caption{Comparison of C-index values on the historical cohort ($\mathcal{H}$) and the combined historical and current cohorts ($\mathcal{H} \cup \mathcal{C}$) from the MIMIC-IV database.}
\label{tab:par2cox-casestudy}
\begin{tabular}{lcc}
\hline
\textbf{Method} \, & $\mathcal{H}$ & $\mathcal{H \cup C}$ \\
\hline

PCA & 0.6122 & 0.6278 \\

CP-WOPT & 0.5588 & 0.5696 \\

DTW-CP & 0.7294 & 0.7444  \\

COPA & 0.7440 & 0.7535 \\

PAR2COX & 0.7610 & 0.7812 \\

\hline
\end{tabular}
\end{table}

\section{Conclusion}\label{sec:conclusion}

Modeling irregular longitudinal EHR data for interpretable phenotype extraction and survival risk prediction remains challenging due to the high dimensionality, sparsity, and irregular follow-up patterns across patients. In this article, we propose PAR2COX, a flexible framework that jointly estimates patient-specific phenotypes and survival risk from irregular longitudinal data. PAR2COX is particularly well-suited for EHR analysis, as it directly handles irregular data without forcing all measurements into a common grid. Furthermore, our framework accommodates both historical and current patients, enabling clinical-decision making and effective treatment allocation in resource-limited settings. In addition, PAR2COX allows the integration of constraints, such as sparsity and temporal smoothness, thereby incorporating prior structural assumptions while enhancing the interpretability of the learned phenotypes.
Numerical experiments and a case study on MIMIC-IV data demonstrate the superiority of PAR2COX for risk stratification, highlighting its ability to recover the underlying phenotypes more effectively than existing approaches. Overall, the proposed framework demonstrates that incorporating survival information directly into the representation learning process leads to more informative patient-specific phenotypes and improved survival prediction from irregular longitudinal data.

Despite the promising results, several limitations remain and provide opportunities for future research.
First, PAR2COX relies on the proportional hazards assumption inherited from the Cox model. Extending the proposed framework to alternative survival models that relax this assumption represents a relevant direction for future research.
Second, although PAR2COX achieves improved risk stratification performance, it is computationally more demanding than conventional two-step approaches due to the joint optimization, as well as the need to select multiple tuning parameters. Improving the computational efficiency of the proposed optimization algorithm constitutes another promising avenue for future work.
Finally, extending PAR2COX to explicitly accommodate missing data would further enhance its applicability in real-world clinical settings.

\appendix

\section*{Disclosure of Interest}
The authors declare no conflicts of interest.

\section*{Data Availability Statement}
The PAR2COX implementation is publicly available at \url{https://github.com/franci2312/PAR2COX}.
The dataset used in the case study is available from PhysioNet \url{https://physionet.org/content/mimiciv/3.1/} to credentialed users; the authors cannot redistribute it.

\section*{Funding}
The PhD scholarship of Mariafrancesca Patalano is funded by the European Union – Next Generation EU, Mission 4, Component 1, CUP C96E23000630001. The work of Kamran Paynabar was partially supported by the Fouts Family Chair Fund.
This research was supported in part through research cyberinfrastructure resources and services provided by the Partnership for an Advanced Computing Environment (PACE) at the Georgia Institute of Technology, Atlanta, Georgia, USA.

\bibliographystyle{apalike}
\bibliography{ref}

@book{kolda2025tensor,
  title={Tensor Decompositions for Data Science},
  author={Kolda, Tamara G. and Ballard, Grey},
  year={2025},
  publisher={SIAM}
}

@article{koldabader2009,
author = {Kolda, Tamara G. and Bader, Brett W.},
title = {Tensor Decompositions and Applications},
journal = {SIAM Review},
volume = {51},
number = {3},
pages = {455-500},
year = {2009},
doi = {10.1137/07070111X},

URL = { 
    
        https://doi.org/10.1137/07070111X
    
    

},
eprint = { 
    
        https://doi.org/10.1137/07070111X
    
    

}
}

@article{katzman2018deepsurv,
  title={DeepSurv: personalized treatment recommender system using a Cox proportional hazards deep neural network},
  author={Katzman, Jared L and Shaham, Uri and Cloninger, Alexander and Bates, Jonathan and Jiang, Tingting and Kluger, Yuval},
  journal={BMC medical research methodology},
  volume={18},
  number={1},
  pages={24},
  year={2018},
  publisher={Springer}
}

@article{ching2018cox,
  title={Cox-nnet: an artificial neural network method for prognosis prediction of high-throughput omics data},
  author={Ching, Travers and Zhu, Xun and Garmire, Lana X},
  journal={PLoS computational biology},
  volume={14},
  number={4},
  pages={e1006076},
  year={2018},
  publisher={Public Library of Science San Francisco, CA USA}
}

@inproceedings{lee2018deephit,
  title={Deephit: A deep learning approach to survival analysis with competing risks},
  author={Lee, Changhee and Zame, William and Yoon, Jinsung and Van Der Schaar, Mihaela},
  booktitle={Proceedings of the AAAI conference on artificial intelligence},
  volume={32},
  number={1},
  year={2018}
}

@article{choi2016retain,
  title={Retain: An interpretable predictive model for healthcare using reverse time attention mechanism},
  author={Choi, Edward and Bahadori, Mohammad Taha and Sun, Jimeng and Kulas, Joshua and Schuetz, Andy and Stewart, Walter},
  journal={Advances in neural information processing systems},
  volume={29},
  year={2016}
}

@article{harshman1972parafac2,
  author    = {Harshman, Richard A.},
  title     = {{PARAFAC2}: Mathematical and Technical Notes},
  journal   = {UCLA Working Papers in Phonetics},
  year      = {1972},
  volume    = {22},
  pages     = {30--44},
  publisher = {University Microfilms},
  address   = {Ann Arbor, Michigan},
  note      = {University Microfilms No. 10,085}
}

@article{cox1972,
 ISSN = {00359246},
 URL = {http://www.jstor.org/stable/2985181},
 author = {D. R. Cox},
 journal = {Journal of the Royal Statistical Society. Series B (Methodological)},
 number = {2},
 pages = {187--220},
 publisher = {[Royal Statistical Society, Oxford University Press]},
 title = {Regression Models and Life-Tables},
 urldate = {2026-05-29},
 volume = {34},
 year = {1972}
}

@article{cox1975,
 ISSN = {00063444, 14643510},
 URL = {http://www.jstor.org/stable/2335362},
 author = {D. R. Cox},
 journal = {Biometrika},
 number = {2},
 pages = {269--276},
 publisher = {[Oxford University Press, Biometrika Trust]},
 title = {Partial Likelihood},
 urldate = {2026-05-29},
 volume = {62},
 year = {1975}
}

@book{klein2003survival,
  author    = {Klein, John P. and Moeschberger, Melvin L.},
  title     = {Survival Analysis: Techniques for Censored and Truncated Data},
  edition   = {2nd},
  publisher = {Springer},
  address   = {New York},
  year      = {2003},
  series    = {Statistics for Biology and Health}
}

@inproceedings{He2019sgranite,
  author    = {He, Huan and Henderson, Jette and Ho, Joyce C.},
  title     = {Distributed Tensor Decomposition for Large Scale Health Analytics},
  booktitle = {Proceedings of the International World-Wide Web Conference},
  year      = {2019},
  pages     = {659--669},
  doi       = {10.1145/3308558.3313548},
  url       = {https://doi.org/10.1145/3308558.3313548}
}

@article{dtwcp,
  author    = {Chi Zhang and Hadi Fanaee-Tork and Magne Thoresen},
  title     = {Feature Extraction from Unequal Length Heterogeneous EHR Time Series via Dynamic Time Warping and Tensor Decomposition},
  journal   = {Data Mining and Knowledge Discovery},
  year      = {2021},
  volume     = {35},
  pages      = {1760--1784},
  doi        = {10.1007/s10618-020-00724-6},
  issn       = {1384-5810},
  publisher  = {Springer}
}

@article{cpwopt,
title = {Scalable tensor factorizations for incomplete data},
journal = {Chemometrics and Intelligent Laboratory Systems},
volume = {106},
number = {1},
pages = {41-56},
year = {2011},
note = {Multiway and Multiset Data Analysis},
issn = {0169-7439},
doi = {https://doi.org/10.1016/j.chemolab.2010.08.004},
url = {https://www.sciencedirect.com/science/article/pii/S0169743910001437},
author = {Evrim Acar and Daniel M. Dunlavy and Tamara G. Kolda and Morten Mørup}
}

@ARTICLE{sspa,
  author={Konyar, Elif and Gahrooei, Mostafa Reisi},
  journal={IEEE Journal of Biomedical and Health Informatics}, 
  title={Semi-Supervised PARAFAC2 Decomposition for Computational Phenotyping Using Electronic Health Records}, 
  year={2025},
  volume={29},
  number={6},
  pages={4415-4425},
  doi={10.1109/JBHI.2025.3530271}}

@article{perros2019,
title = {Temporal phenotyping of medically complex children via PARAFAC2 tensor factorization},
journal = {Journal of Biomedical Informatics},
volume = {93},
pages = {103125},
year = {2019},
issn = {1532-0464},
doi = {https://doi.org/10.1016/j.jbi.2019.103125},
url = {https://www.sciencedirect.com/science/article/pii/S1532046419300437},
author = {Ioakeim Perros and Evangelos E. Papalexakis and Richard Vuduc and Elizabeth Searles and Jimeng Sun}
}

@article{zhao2019,
title = {Detecting time-evolving phenotypic topics via tensor factorization on electronic health records: Cardiovascular disease case study},
journal = {Journal of Biomedical Informatics},
volume = {98},
pages = {103270},
year = {2019},
issn = {1532-0464},
doi = {https://doi.org/10.1016/j.jbi.2019.103270},
url = {https://www.sciencedirect.com/science/article/pii/S1532046419301893},
author = {Juan Zhao and Yun Zhang and David J. Schlueter and Patrick Wu and Vern {Eric Kerchberger} and S. {Trent Rosenbloom} and Quinn S. Wells and QiPing Feng and Joshua C. Denny and Wei-Qi Wei}
}

@inproceedings{copa,
author = {Afshar, Ardavan and Perros, Ioakeim and Papalexakis, Evangelos E. and Searles, Elizabeth and Ho, Joyce and Sun, Jimeng},
title = {COPA: Constrained PARAFAC2 for Sparse \& Large Datasets},
year = {2018},
isbn = {9781450360142},
publisher = {Association for Computing Machinery},
address = {New York, NY, USA},
url = {https://doi.org/10.1145/3269206.3271775},
doi = {10.1145/3269206.3271775},
booktitle = {Proceedings of the 27th ACM International Conference on Information and Knowledge Management},
pages = {793–802},
numpages = {10},
location = {Torino, Italy},
series = {CIKM '18}
}

@inproceedings{spartan,
author = {Perros, Ioakeim and Papalexakis, Evangelos E. and Wang, Fei and Vuduc, Richard and Searles, Elizabeth and Thompson, Michael and Sun, Jimeng},
title = {SPARTan: Scalable PARAFAC2 for Large \& Sparse Data},
year = {2017},
isbn = {9781450348874},
publisher = {Association for Computing Machinery},
address = {New York, NY, USA},
url = {https://doi.org/10.1145/3097983.3098014},
doi = {10.1145/3097983.3098014},
booktitle = {Proceedings of the 23rd ACM SIGKDD International Conference on Knowledge Discovery and Data Mining},
pages = {375–384},
numpages = {10},
location = {Halifax, NS, Canada},
series = {KDD '17}
}

@INPROCEEDINGS{atom,
  author={Jang, Jun-Gi and Lee, Jeongyoung and Park, Jiwon and Kang, U},
  booktitle={2022 IEEE International Conference on Big Data (Big Data)}, 
  title={Accurate PARAFAC2 Decomposition for Temporal Irregular Tensors with Missing Values}, 
  year={2022},
  volume={},
  number={},
  pages={982-991},
  doi={10.1109/BigData55660.2022.10020667}}

@inproceedings{logpar,
author = {Yin, Kejing and Afshar, Ardavan and Ho, Joyce C. and Cheung, William K. and Zhang, Chao and Sun, Jimeng},
title = {LogPar: Logistic PARAFAC2 Factorization for Temporal Binary Data with Missing Values},
year = {2020},
isbn = {9781450379984},
publisher = {Association for Computing Machinery},
address = {New York, NY, USA},
url = {https://doi.org/10.1145/3394486.3403213},
doi = {10.1145/3394486.3403213},
booktitle = {Proceedings of the 26th ACM SIGKDD International Conference on Knowledge Discovery \& Data Mining},
pages = {1625–1635},
numpages = {11},
location = {Virtual Event, CA, USA},
series = {KDD '20}
}

@inproceedings{repair,
author = {Ren, Yifei and Lou, Jian and Xiong, Li and Ho, Joyce C.},
title = {Robust Irregular Tensor Factorization and Completion for Temporal Health Data Analysis},
year = {2020},
isbn = {9781450368599},
publisher = {Association for Computing Machinery},
address = {New York, NY, USA},
url = {https://doi.org/10.1145/3340531.3411982},
doi = {10.1145/3340531.3411982},
booktitle = {Proceedings of the 29th ACM International Conference on Information \& Knowledge Management},
pages = {1295–1304},
numpages = {10},
location = {Virtual Event, Ireland},
series = {CIKM '20}
}

@inproceedings{marble,
author = {Ho, Joyce C. and Ghosh, Joydeep and Sun, Jimeng},
title = {Marble: high-throughput phenotyping from electronic health records via sparse nonnegative tensor factorization},
year = {2014},
isbn = {9781450329569},
publisher = {Association for Computing Machinery},
address = {New York, NY, USA},
url = {https://doi.org/10.1145/2623330.2623658},
doi = {10.1145/2623330.2623658},
booktitle = {Proceedings of the 20th ACM SIGKDD International Conference on Knowledge Discovery and Data Mining},
pages = {115–124},
numpages = {10},
location = {New York, New York, USA},
series = {KDD '14}
}

@book{jolliffe,
author = {Jolliffe, I.},
address = {New York},
booktitle = {Principal component analysis},
edition = {2. ed },
isbn = {0387954422},
language = {eng},
publisher = {Springer},
series = {Springer series in statistics},
title = {Principal component analysis},
year = {2002},
}

@article{nmf,
  author  = {Lee, Daniel D. and Seung, H. Sebastian},
  title   = {Learning the Parts of Objects by Non-Negative Matrix Factorization},
  journal = {Nature},
  volume   = {401},
  number   = {6755},
  pages    = {788--791},
  year     = {1999},
  doi      = {10.1038/44565}
}

@article{singer2016third,
  title={The third international consensus definitions for sepsis and septic shock (Sepsis-3)},
  author={Singer, Mervyn and Deutschman, Clifford S and Seymour, Christopher Warren and Shankar-Hari, Manu and Annane, Djillali and Bauer, Michael and Bellomo, Rinaldo and Bernard, Gordon R and Chiche, Jean-Daniel and Coopersmith, Craig M and others},
  journal={Jama},
  volume={315},
  number={8},
  pages={801--810},
  year={2016}
}

@article{evans2021surviving,
  title={Surviving sepsis campaign: international guidelines for management of sepsis and septic shock 2021},
  author={Evans, Laura and Rhodes, Andrew and Alhazzani, Waleed and Antonelli, Massimo and Coopersmith, Craig M and French, Craig and Machado, Fl{\'a}via R and Mcintyre, Lauralyn and Ostermann, Marlies and Prescott, Hallie C and others},
  journal={Critical care medicine},
  volume={49},
  number={11},
  pages={e1063--e1143},
  year={2021},
  publisher={LWW}
}

@misc{nmfcox,
      title={Low-Rank Reorganization via Proportional Hazards Non-negative Matrix Factorization Unveils Survival Associated Gene Clusters}, 
      author={Zhi Huang and Paul Salama and Wei Shao and Jie Zhang and Kun Huang},
      year={2020},
      eprint={2008.03776},
      archivePrefix={arXiv},
      primaryClass={q-bio.QM},
      url={https://arxiv.org/abs/2008.03776}, 
}

@article{shivade2014review,
  title={A review of approaches to identifying patient phenotype cohorts using electronic health records},
  author={Shivade, Chaitanya and Raghavan, Preethi and Fosler-Lussier, Eric and Embi, Peter J and Elhadad, Noemie and Johnson, Stephen B and Lai, Albert M},
  journal={Journal of the American Medical Informatics Association},
  volume={21},
  number={2},
  pages={221--230},
  year={2014},
  publisher={BMJ Publishing Group}
}

@article{Kiers1999,
  author  = {Kiers, Henk A. L. and Ten Berge, Jos M. F. and Bro, Rasmus},
  title   = {PARAFAC2---Part I. A Direct Fitting Algorithm for the PARAFAC2 Model},
  journal = {Journal of Chemometrics},
  year    = {1999},
  volume  = {13},
  number  = {3--4},
  pages   = {275--294},
  doi     = {10.1002/(SICI)1099-128X(199905/08)13:3/4<275::AID-CEM543>3.3.CO;2-2}
}

@article{chen2025,
author = {Chen, Chin-Chun and Chang, Sheng-Mao and Lin, Peng-Chan and Su, Pei-Fang},
title = {Statistical Inference for Tensor-Based Covariates in Cox Regression Modeling in Integrated Genome Studies},
journal = {Statistics in Medicine},
volume = {44},
number = {13-14},
pages = {e70145},
doi = {https://doi.org/10.1002/sim.70145},
url = {https://onlinelibrary.wiley.com/doi/abs/10.1002/sim.70145},
eprint = {https://onlinelibrary.wiley.com/doi/pdf/10.1002/sim.70145},
year = {2025}
}

@article{huangaoadmm,
author = {Huang, Kejun and Sidiropoulos, Nicholas D. and Liavas, Athanasios P.},
title = {A Flexible and Efficient Algorithmic Framework for Constrained Matrix and Tensor Factorization},
year = {2016},
issue_date = {Oct.1, 2016},
publisher = {IEEE Press},
volume = {64},
number = {19},
issn = {1053-587X},
url = {https://doi.org/10.1109/TSP.2016.2576427},
doi = {10.1109/TSP.2016.2576427},
journal = {Trans. Sig. Proc.},
month = oct,
pages = {5052–5065},
numpages = {14}
}

@article{hripcsak2013next,
  title={Next-generation phenotyping of electronic health records},
  author={Hripcsak, George and Albers, David J},
  journal={Journal of the American Medical Informatics Association},
  volume={20},
  number={1},
  pages={117--121},
  year={2013},
  publisher={BMJ Group}
}

@article{shickel2017deep,
  title={Deep EHR: a survey of recent advances in deep learning techniques for electronic health record (EHR) analysis},
  author={Shickel, Benjamin and Tighe, Patrick James and Bihorac, Azra and Rashidi, Parisa},
  journal={IEEE journal of biomedical and health informatics},
  volume={22},
  number={5},
  pages={1589--1604},
  year={2017},
  publisher={IEEE}
}

@article{jensen2012mining,
  title={Mining electronic health records: towards better research applications and clinical care},
  author={Jensen, Peter B and Jensen, Lars J and Brunak, S{\o}ren},
  journal={Nature Reviews Genetics},
  volume={13},
  number={6},
  pages={395--405},
  year={2012},
  publisher={Nature Publishing Group UK London}
}

@article{Jia2015GeneRanking,
  author  = {Jia, Zhihua and Zhang, Xiaohui and Guan, Nan and Bo, Xin and Barnes, Michael R. and Luo, Zhihua},
  title   = {Gene Ranking of RNA-Seq Data via Discriminant Non-Negative Matrix Factorization},
  journal = {PLOS ONE},
  volume   = {10},
  number   = {9},
  pages    = {e0137782},
  year     = {2015},
  doi      = {10.1371/journal.pone.0137782}
}

@article{vandermaaten08,
  author  = {Laurens van der Maaten and Geoffrey Hinton},
  title   = {Visualizing Data using t-SNE},
  journal = {Journal of Machine Learning Research},
  year    = {2008},
  volume  = {9},
  number  = {86},
  pages   = {2579--2605},
  url     = {http://jmlr.org/papers/v9/vandermaaten08a.html}
}

@article{harrell1982,
  title     = {Evaluating the yield of medical tests},
  author    = {Harrell, Frank E. and Califf, Robert M. and Pryor, David B. and Lee, Kerry L. and Rosati, Robert A.},
  journal   = {JAMA},
  volume    = {247},
  number    = {18},
  pages     = {2543--2546},
  year      = {1982},
  month     = {May},
  doi       = {10.1001/jama.1982.03320430047030},
  pmid      = {7069920}
}

@article{bender2005generating,
  title={Generating survival times to simulate Cox proportional hazards models},
  author={Bender, Ralf and Augustin, Thomas and Blettner, Maria},
  journal={Statistics in medicine},
  volume={24},
  number={11},
  pages={1713--1723},
  year={2005},
  publisher={Wiley Online Library}
}

@article{Rudd2020Sepsis,
  author    = {Kristina E. Rudd and Sarah Charlotte Johnson and Kareha M. Agesa and Katya Anne Shackelford and Derrick Tsoi and Daniel Rhodes Kievlan and Danny V. Colombara and Kevin S. Ikuta and Niranjan Kissoon and Simon Finfer and Carolin Fleischmann-Struzek and Flavia R. Machado and Konrad K. Reinhart and Kathryn Rowan and Christopher W. Seymour and R. Scott Watson and T. Eoin West and Fatima Marinho and Simon I. Hay and Rafael Lozano and Alan D. Lopez and Derek C. Angus and Christopher J. L. Murray and Mohsen Naghavi},
  title      = {Global, Regional, and National Sepsis Incidence and Mortality, 1990--2017: Analysis for the Global Burden of Disease Study},
  journal    = {The Lancet},
  year       = {2020},
  volume     = {395},
  number     = {10219},
  pages      = {200--211},
  doi        = {10.1016/S0140-6736(19)32989-7},
  issn       = {0140-6736},
  publisher  = {Elsevier}
}

@misc{Johnson2024MIMICIV,
  author       = {Johnson, Alistair and Bulgarelli, Lucas and Pollard, Tom and Gow, Brian and Moody, Benjamin and Horng, Steven and Celi, Leo Anthony and Mark, Roger},
  title        = {{MIMIC-IV} (version 3.1)},
  year         = {2024},
  publisher    = {PhysioNet},
  doi          = {10.13026/kpb9-mt58},
  note         = {RRID:SCR\_007345}
}

@article{heagerty2000time,
  title={Time-Dependent ROC Curves for Censored Survival Data and a Diagnostic Marker},
  author={Heagerty, Patrick J. and Lumley, Thomas and Pepe, Margaret S.},
  journal={Biometrics},
  volume={56},
  number={2},
  pages={337--344},
  year={2000},
  publisher={Wiley},
  doi={10.1111/j.0006-341X.2000.00337.x}
}

@article{Vincent1996SOFA,
  author = {Vincent, Jean-Louis and Moreno, Rui and Takala, Jukka and Willatts, Simon and De Mendon{\c{c}}a, Arnaldo and Bruining, Hugo and Reinhart, Konrad K. and Suter, Peter M. and Thijs, L. G.},
  title = {The {SOFA} ({S}epsis-related {O}rgan {F}ailure {A}ssessment) score to describe organ dysfunction/failure},
  journal = {Intensive Care Medicine},
  year = {1996},
  volume = {22},
  number = {7},
  pages = {707--710},
  doi = {10.1007/BF01709751}
}

@article{Johnson2023paper,
  author  = {Johnson, Alistair E. W. and Bulgarelli, Lucas and Shen, Li-wei and others},
  title   = {{MIMIC-IV}, a Freely Accessible Electronic Health Record Dataset},
  journal = {Scientific Data},
  year    = {2023},
  volume  = {10},
  number  = {1},
  pages   = {1},
  doi     = {10.1038/s41597-022-01899-x}
}

@inproceedings{Chao2018ICUMortality,
  author    = {Chao, Gang and Mao, Chengsheng and Wang, Fei and Zhao, Yuan and Luo, Yonghui},
  title     = {Supervised Nonnegative Matrix Factorization to Predict ICU Mortality Risk},
  booktitle = {Proceedings of the IEEE International Conference on Bioinformatics and Biomedicine (BIBM)},
  pages      = {1189--1194},
  year       = {2018},
  doi        = {10.1109/BIBM.2018.8621403}
}

@inproceedings{rubik,
author = {Wang, Yichen and Chen, Robert and Ghosh, Joydeep and Denny, Joshua C. and Kho, Abel and Chen, You and Malin, Bradley A. and Sun, Jimeng},
title = {Rubik: Knowledge Guided Tensor Factorization and Completion for Health Data Analytics},
year = {2015},
isbn = {9781450336642},
publisher = {Association for Computing Machinery},
address = {New York, NY, USA},
url = {https://doi.org/10.1145/2783258.2783395},
doi = {10.1145/2783258.2783395},
booktitle = {Proceedings of the 21th ACM SIGKDD International Conference on Knowledge Discovery and Data Mining},
pages = {1265–1274},
numpages = {10},
location = {Sydney, NSW, Australia},
series = {KDD '15}
}

@article{
brunet2004,
author = {Jean-Philippe Brunet  and Pablo Tamayo  and Todd R. Golub  and Jill P. Mesirov },
title = {Metagenes and molecular pattern discovery using matrix factorization},
journal = {Proceedings of the National Academy of Sciences},
volume = {101},
number = {12},
pages = {4164-4169},
year = {2004},
doi = {10.1073/pnas.0308531101},
URL = {https://www.pnas.org/doi/abs/10.1073/pnas.0308531101},
eprint = {https://www.pnas.org/doi/pdf/10.1073/pnas.0308531101}}

@article{Shen2006pca,
  author  = {Shen, Yu-Jun and Huang, Shih-Gang},
  title   = {Improve Survival Prediction Using Principal Components of Gene Expression Data},
  journal = {Genomics, Proteomics \& Bioinformatics},
  volume   = {4},
  number   = {2},
  pages    = {110--119},
  year     = {2006},
  doi      = {10.1016/S1672-0229(06)60022-3}
}

\end{document}